\documentclass[a4paper,11pt]{article}
\pdfoutput=1

\usepackage[utf8]{inputenc}
\usepackage[english, italian]{babel}
\usepackage[T1]{fontenc}

\usepackage[titletoc,toc,title]{appendix}

\usepackage{graphicx}
\usepackage{amsmath}
\usepackage{amsfonts}
\usepackage{float}

\usepackage{booktabs}
\usepackage{array}
\usepackage{paralist}
\usepackage{verbatim}
\usepackage{subfig}
\usepackage{multirow}
\usepackage{empheq}

\usepackage{feynmf}
\usepackage{rotating}

\include{caption}
\include{subfig}

\usepackage{adjustbox}

\usepackage{color} 
\usepackage{cite}

\usepackage{hyperref}

\renewcommand{\epsilon}{\varepsilon}
\newcommand{\ep}{\varepsilon}
\renewcommand{\theta}{\vartheta}
\renewcommand{\rho}{\varrho}
\renewcommand{\phi}{\varphi}

	{\left\lbrace\begin{array}{@{}l@{}}}%
	{\end{array}\right.}

\newcommand{\imineq}[2]{\vcenter{\hbox{\includegraphics[height=#2ex]{#1}}}}

\newsavebox{\overlongequation}
\newenvironment{bigeq}
 {\begin{displaymath}\begin{lrbox}{\overlongequation}$\displaystyle}
 {$\end{lrbox}\makebox[0pt]{\usebox{\overlongequation}}\end{displaymath}}

\begin{document}
\selectlanguage{english}

\setlength{\unitlength}{1.3cm} 
\begin{titlepage}
\vspace*{-1cm}
\begin{flushright}
TTP16-045
\end{flushright}                                
\vskip 3.5cm
\begin{center}
\boldmath
 
{\Large\bf Two-loop electroweak  corrections to Higgs-gluon couplings to higher orders in the dimensional regularization parameter\\[3mm] }
\unboldmath
\vskip 1.cm
{\large Marco Bonetti}
\footnote{{\tt marco.bonetti@kit.edu}},
{\large Kirill Melnikov}
\footnote{{\tt kirill.melnikov@kit.edu}} and
{\large Lorenzo Tancredi}
\footnote{{\tt lorenzo.tancredi@kit.edu}} 
\vskip .7cm
{\it Institute for Theoretical Particle Physics, KIT, 76128 Karlsruhe, Germany } 
\end{center}
\vskip 2.6cm

\begin{abstract}
We compute the two-loop electroweak correction to the production of the Higgs boson in gluon fusion to higher orders in the dimensional-regularization parameter $\epsilon = (d-4)/2$. 
We employ  the method of differential equations augmented by the choice of a canonical basis to compute the relevant integrals and express them in terms of Goncharov polylogarithms. Our calculation provides useful results for the computation of the NLO mixed QCD-electroweak corrections to $gg \to H$ and establishes the necessary framework towards the calculation of the missing three-loop virtual corrections.

\end{abstract}
\vfill
\end{titlepage}                                                                
\newpage


\section{Introduction}\setcounter{equation}{0} 
\numberwithin{equation}{section}
The recent discovery of the  Higgs boson and the non-observation of any New Physics at the LHC 
establishes the validity of the Standard Model  as the low-energy effective theory of Nature. 
At the same time, the apparent inability of the Standard Model to explain several experimental 
facts makes the need for physics beyond the Standard Model (BSM) as strong as ever. Searching for 
clues about BSM physics is in the focus of contemporary particle physics. 
The Higgs boson is bound to play an important role in this endeavor. Indeed, 
the Higgs mechanism in the Standard Model is very simplistic and rather \textit{ad hoc}. 
At the same time, there are many extensions of the Standard Model where 
the Higgs boson  is the only particle that is sensitive to rich physics beyond it. 
More generally, if  the Higgs boson is responsible 
for generating masses not only of the Standard Model but also of some BSM particles, which 
appears to be necessary for protecting the Higgs boson mass from large radiative corrections, 
these new particles may affect the 
couplings of the Higgs boson to gauge bosons and fermions through radiative corrections. 
A percent  modification
of the Higgs couplings  is a generic consequence of physics beyond the Standard Model at the energy 
scale of about $1~{\rm TeV}$.  Therefore, measurement of  the Higgs bosons couplings to Standard Model 
particles to this level of precision is an important goal of the LHC physics program. 

The major  production channel of Higgs bosons at the LHC is gluon fusion. 
The recent computation of the three-loop QCD corrections to $\sigma_{gg \to H}$ significantly reduces theoretical 
uncertainty in the predicted cross section. According to Ref.~\cite{Anastasiou:2016cez},  the 
theory uncertainty in $\sigma_{gg \to H}$ is close to $5\%$ and the uncertainties related to 
imprecise knowledge of parton distribution functions and the strong coupling constant 
are  close to $4\%$.  The theoretical uncertainty has several sources such as  the residual 
scale dependence of the three-loop QCD result, imperfect 
knowledge of the bottom quark contribution to $gg \to H$ and the mixed three-loop QCD-electroweak corrections 
which are known in the unphysical limit  $m_{Z,W} \gg m_H$ \cite{Anastasiou:2008tj}.  Each of these sources 
contributes similar amount to the final uncertainty which implies that a  better understanding of all of them 
is required for reducing the uncertainty to $\sim 1$-$2$\%.

In this paper we focus on the computation of the two-loop electroweak correction  to the production 
of the Higgs boson in gluon fusion. This  contribution arises because gluons couple to electroweak vector 
bosons $W$ and $Z$ through a quark loop; a subsequent  fusion of the electroweak bosons to the Higgs boson 
gives rise to electroweak-mediated $ggH$ coupling.  The quark loop receives contributions from both light 
and heavy quarks but the relatively small mass of the Higgs boson leads to a strong dominance of the 
light quark contributions.\footnote{More precisely, about $95\%$ of the full electroweak 
contribution to $ggH$ is due to the light quark loops.}

The electroweak  contributions to $ggH$ have been evaluated analytically at leading (two-loop) order 
in Refs.~\cite{Aglietti:2006yd, Aglietti:2004nj, Aglietti:2004ki,Actis:2008ts}.  Since the QCD corrections increase 
the leading, top-quark mediated,  contribution to $gg \to H$ by almost a factor two, it is essential to understand if 
a similar enhancement is present  in case of  the electroweak corrections 
to $gg \to H$.  To clarify this issue, we need  a computation of QCD corrections to the 
electroweak contribution to $ggH$. However, since the electroweak contribution starts at two loops, 
calculation of NLO QCD corrections requires dealing with three-loop diagrams with massive 
internal lines.  Given the complexity  of the required  computation, one can try to simplify 
it by considering different kinematic limits: mixed QCD-electroweak corrections 
in   the unphysical limit of a vanishingly small Higgs boson mass 
$m_{Z,W} \gg m_H$ were calculated in  Ref.~\cite{Anastasiou:2008tj}.

However, recent progress in  the theoretical understanding  of QCD effects in $gg~\to~H$ and 
continuous developments in the technology of multi-loop computations make  it worthwhile and interesting 
to attempt an exact computation of the NLO QCD corrections to the electroweak contribution to 
$ggH$.  In this paper, we make  an important step in this direction by setting up a 
modern calculational framework for this problem 
that employs canonical bases for master integrals and differential equations, and
computing the two-loop electroweak 
contribution to $ggH$ to higher orders in the dimensional regularization parameter $\epsilon=(4-d)/2$. 
The knowledge of the two-loop amplitude to higher orders in $\ep$ is necessary for subtracting 
infrared and collinear singularities from the electroweak contributions to the $gg \to Hg$  
inelastic process or, alternatively, for extracting the relevant finite remainder,  defined by  
the Catani formula \cite{Catani:1998bh},  from the three-loop mixed QCD-electroweak contribution to $ggH$ amplitude. 

Specifically, we derive the two-loop electroweak correction to $gg \to H$ through 
${\cal O}(\ep^2)$  and show that only GPLs up to weight five appear in this amplitude. 
We perform our calculation using the method of differential 
equations~\cite{Kotikov:1990kg,Remiddi:1997ny,Gehrmann:1999as},
augmented by the choice of a \emph{canonical basis} of master integrals, introduced in Ref.~\cite{Henn:2013pwa}.\footnote{An alternative way to construct and solve differential equations has been investigated in Ref.~\cite{Papadopoulos:2014lla}.} 
A canonical basis of master integrals is presented and the 
master integrals are calculated  in terms of  
Goncharov's multiple polylogarithms (GPLs) \cite{Remiddi:1999ew, Goncharov, Vollinga:2004sn}.
In order to fix analytically all boundary conditions we make extensive use of the large mass expansion
the PSLQ algorithm.
This allows us to derive the expansion for the master integrals in the dimensional regularization parameter 
$\epsilon$ through weight six.  
From our calculation we can easily reproduce the ${\cal O}(\ep^0)$
results which have been known for a long time in the literature~\cite{Aglietti:2004ki}.

The paper is organized as follows. We introduce the  notation and discuss the structure of the scattering 
amplitude $gg \to H$ in Section~\ref{tensdec}. We describe the master integrals  and the differential equations 
in  Section~\ref{MI}.    We  explain how the boundary conditions 
can be  fixed using the large mass expansion  and 
outline the analytic continuation of GPLs, required 
to obtain results in physical kinematics, in Section~\ref{BC}.
The $gg\to H$ finite part of the amplitude is given in Section~\ref{ampli}.
We provide constants of integration for the master integrals up to weight six in appendix~\ref{AB}. The explicit expressions for the master integrals up to this weight, and the $gg\to H$ amplitude through $\mathcal{O}\left(\epsilon^2\right)$ are available in the ancillary file.

\section{Feynman diagrams and master integrals}
\label{tensdec}\setcounter{equation}{0} 
\numberwithin{equation}{section}

\begin{figure}[h]
      \centering
      \subfloat
      {\includegraphics[height=0.12\textheight]{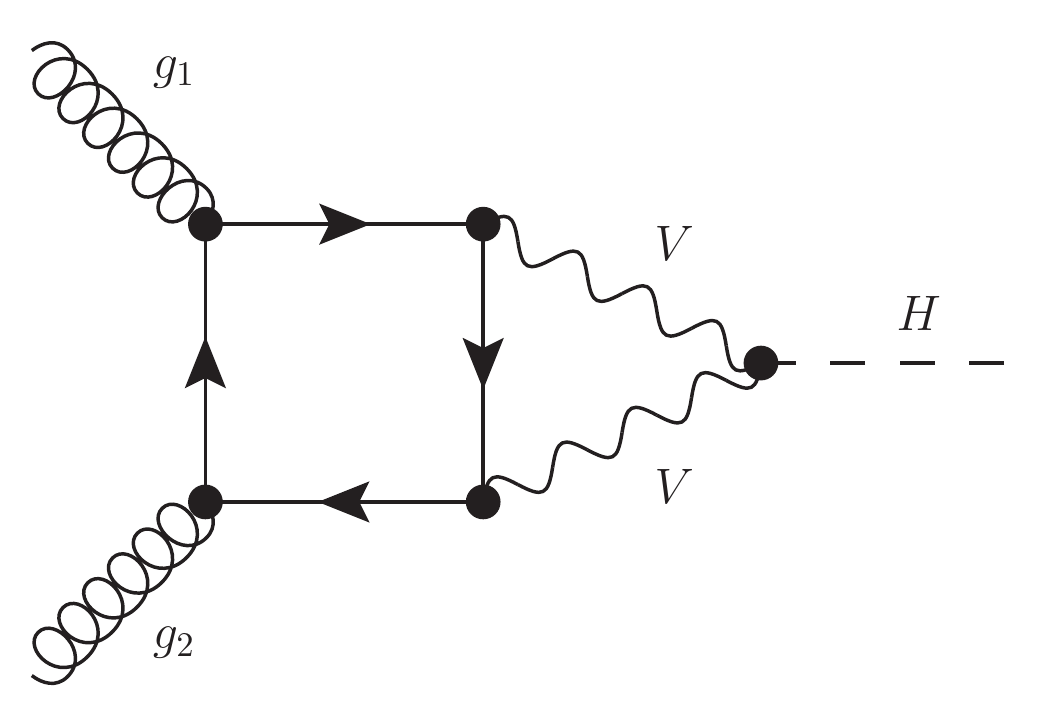}}
      \qquad
      \subfloat
      {\includegraphics[height=0.12\textheight]{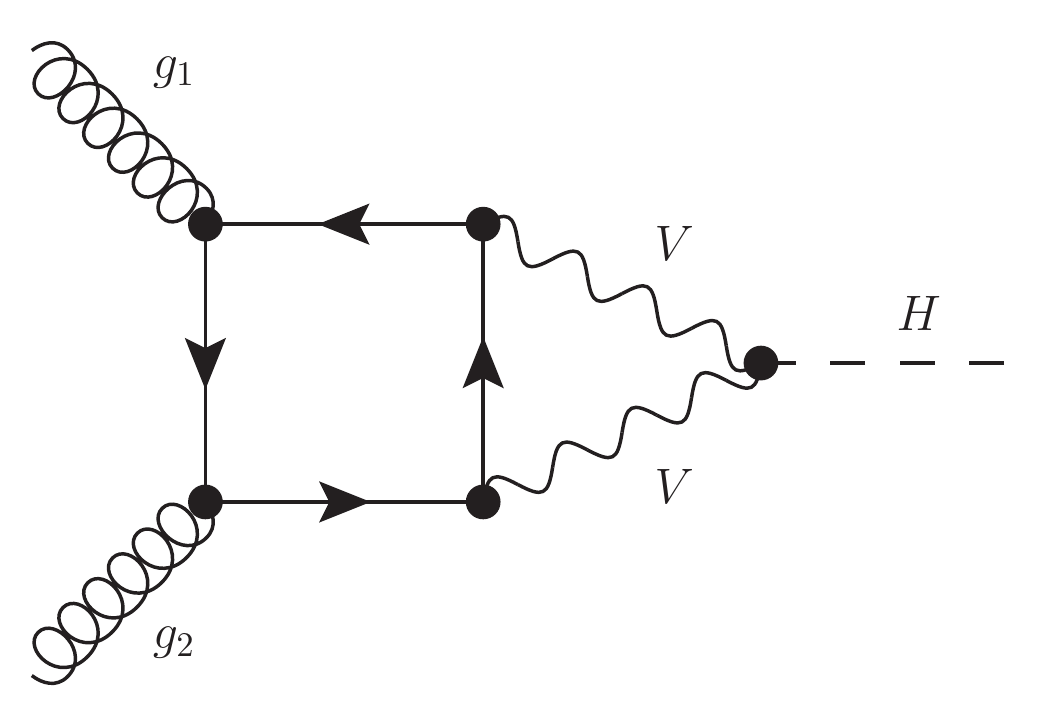}}
      \qquad
      \subfloat
      {\includegraphics[height=0.12\textheight]{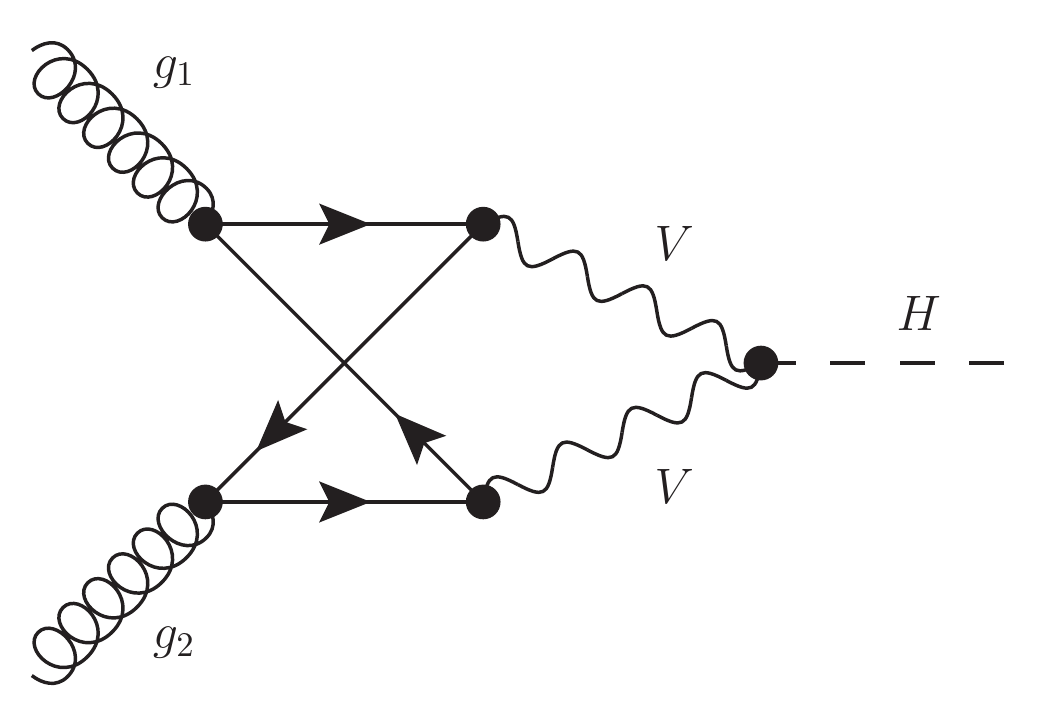}}
      \caption{The two-loop light-quark electroweak contributions to $gg\to H$. 
$V$ stands for $W^\pm, Z$ and the fermionic lines represent different quarks, depending on the electroweak boson $V$.}
      \label{f1}
\end{figure}

We consider the electroweak correction 
 to  the $gg \to H$ amplitude mediated by a light-quark loop.  The relevant contributions are 
shown in Figure~\ref{f1}. The fermionic lines represent up, down, strange and charm quarks, that are taken to be massless.\footnote{Bottom quarks require a special treatment, together with top quarks.} 
The incoming gluons $g_1$ and $g_2$ are on-shell and carry  momenta $p_1$ and $p_2$, with color indices $c_1$ and $c_2$ and polarizations $\epsilon_{\lambda_1}(\mathbf{p}_1)$ and $\epsilon_{\lambda_2}(\mathbf{p}_2)$, respectively. The momentum of the 
Higgs boson is taken to be $p_3 = p_1+p_2$ with $p_3^2 = m_H^2 = s$. 

Thanks to gauge-invariance and parity constraints,
the $gg \to H$ scattering amplitude is expressed in terms of a single form factor
\begin{equation}
 \mathcal{M}^{c_1c_2}_{\lambda_1\lambda_2}=\mathcal{F}\left(s,m_W^2,m_Z^2\right)
\left[\eta_{\mu\nu}-\frac{p_{2\mu}p_{1\nu}}{p_1\cdot p_2}\right]\delta^{c_1c_2}\epsilon_{\lambda_1}^\mu(\mathbf{p}_1)\epsilon_{\lambda_2}^\nu
(\mathbf{p}_2).
\end{equation}
It is possible to extract the form factor $\mathcal{F}$ by contracting  $\mathcal{M}_{\lambda_1\lambda_2}^{c_1c_2}$ with the projection operator
\begin{equation}
 \mathbb{P}_{c_1c_2}^{\lambda_1\lambda_2}=\epsilon^{*\lambda_1}_\mu(\mathbf{p}_1)
\epsilon^{*\lambda_2}_\nu(\mathbf{p}_2)\frac{\delta_{c_1c_2}}{N_c^2-1}\frac{1}{d-2}\left[\eta^{\mu\nu}-\frac{p_1^\mu p_2^\nu+p_1^\nu p_2^\mu}{p_1\cdot p_2}\right].
\end{equation}
We find
\begin{equation}
 \mathcal{F}\left(s,m_W^2,m_Z^2\right)=\sum_{\lambda_1,\lambda_2,c_1,c_2}\mathbb{P}_{c_1,c_2}^{\lambda_1\lambda_2}\mathcal{M}_{\lambda_1\lambda_2}^
{c_1,c_2}.
\end{equation}

\begin{figure}[h]
      \centering
      \subfloat
      {\includegraphics[width=0.35\textwidth]{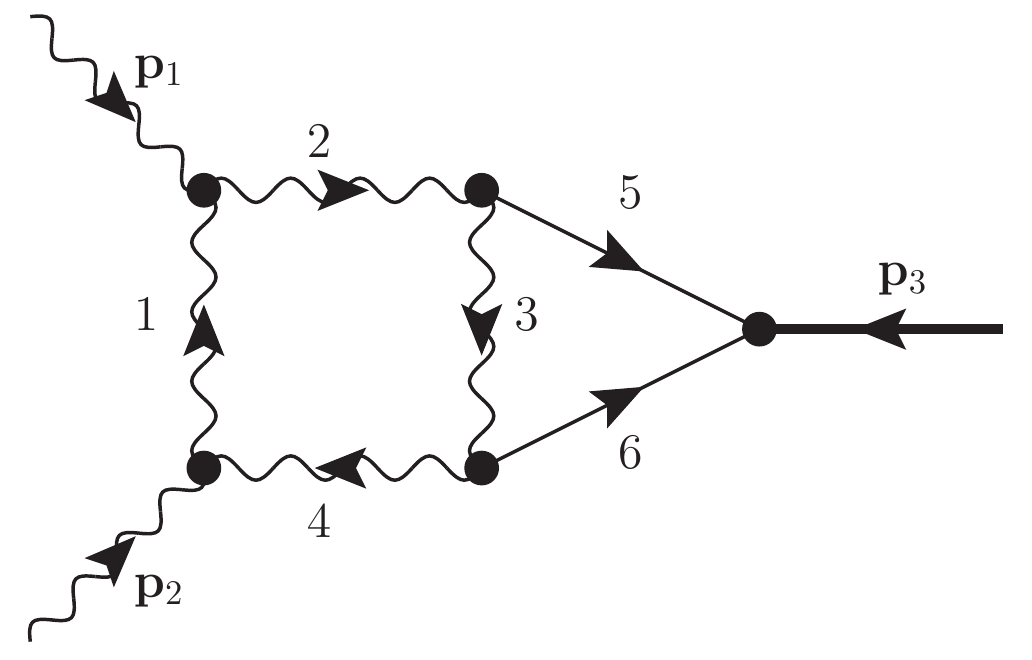}}
      \quad
      \subfloat
      {\includegraphics[width=0.35\textwidth]{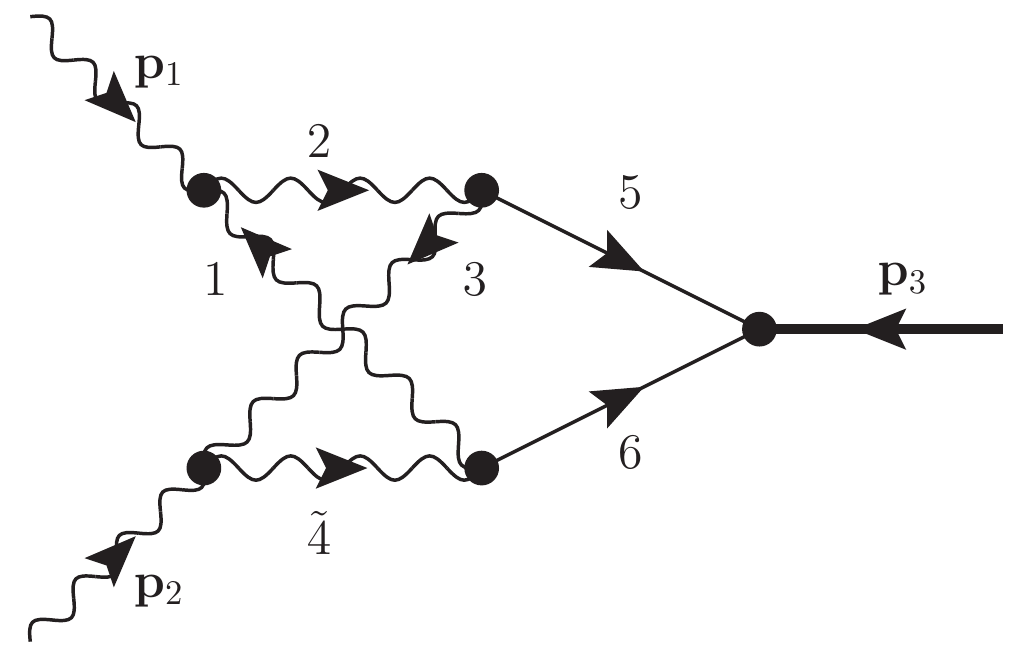}}
      \caption{Planar (left) and non-planar (right) topologies.  See text for momenta assignments 
and propagator labels.}
      \label{f2}
\end{figure}

The form factor $\mathcal{F}$ is a linear combination of integrals which depend on the scalar products between loop and external momenta, and on the scalar products of loop momenta between themselves. All the integrals in $\mathcal{F}$ are obtained starting from the two topologies shown in Figure~\ref{f2}. At variance with Feynman diagrams in Figure~\ref{f1}, when we consider topologies and master integrals, we use wavy (solid) lines to denote massless (massive) propagators, respectively.
We take all momenta to be incoming, i.e. 
\begin{equation}
p_3 = -p_1 - p_2,\;\;\; p_{1,2}^2 = 0,\;\;\;\;  p_3^3 = s = m_H^2.
\end{equation}

The planar and non-planar integrals are parametrized as 
\begin{align} 
& I_\textup{P}(a_1,a_2,a_3,a_4,a_5,a_6,a_7)&= \int \frac{{\rm d}^dk_1 {\rm d}^d k_2}{[\mathrm{i} \mathrm{\pi}^2 \Gamma(1+\ep)]^2}
\frac{1}{ [1]^{a_1} [2]^{a_2} [3]^{a_3} [4]^{a_4} [5]^{a_5} [6]^{a_6} [7]^{a_7}    },
\\
& I_\textup{NP}(a_1,a_2,a_3,a_{\tilde{4}},a_5,a_6,a_7)&= \int \frac{{\rm d}^dk_1 {\rm d}^d k_2}{[\mathrm{i} \mathrm{\pi}^2 \Gamma(1+\ep)]^2}
\frac{1}{ [1]^{a_1} [2]^{a_2} [3]^{a_3} [\tilde 4]^{a_{\tilde{4}}} [5]^{a_5} [6]^{a_6} [7]^{a_7}    },
\label{topo1}
\end{align}
where 
\begin{equation}
\label{topo}
\begin{split}
& [1] = k_1^2,\quad [2] = (k_1+p_1)^2,\quad[3] = k_2^2,\quad[4] = (k_1-p_2)^2,\quad
[\tilde{4}] = (k_2+p_2)^2,\\
& [5] = (k_1-k_2+p_1)^2-M^2, \quad [6] = (k_2-k_1+p_2)^2-M^2, \quad
[7] = (k_1+k_2)^2. 
\end{split}
\end{equation}
In both cases, the propagator $[7]$ is auxiliary;  it is only needed for the parametrization of tensor integrals 
with (otherwise) irreducible numerators.
Both planar and non-planar  integrals are analytic functions in the complex plane of the 
variable $s$ with the cut along the real axis starting at  $s = 0$. This discontinuity 
corresponds to massless intermediate states in Feynman diagrams.  
At $s \ge  4M^2$, it also becomes possible to produce pairs of vector bosons on the mass shell; 
this leads to additional contributions to the discontinuities of $I_{\textup{P,NP}}$. 
We use the program \texttt{Reduze2}~\cite{vonManteuffel:2012np} to express 
all integrals 
that appear  in the evaluation of $gg \to H$ amplitude 
through master integrals (MIs). We also use the integration-by-parts reduction identities 
to derive the differential equations in $s$ and $M^2$ satisfied by the master integrals.

\section{Differential equations}
\label{MI}\setcounter{equation}{0} 
\numberwithin{equation}{section}

We denote a vector of master integrals by ${\bf I}$, 
a set of kinematic variables by ${\bf x} \in \{s,M^2 \}$,  
and write the differential equations as
\begin{equation}
 \frac{\partial\mathbf{I}(\mathbf{x},\epsilon)}{\partial\mathbf{x}_i}
={\cal A}_i(\mathbf{x},\epsilon)\mathbf{I}(\mathbf{x},\epsilon).
\end{equation}
It was conjectured in Ref.~\cite{Henn:2013pwa} that in many physically relevant cases 
a {\it canonical} basis of master integrals ${\bf I'}$ 
exists with the property that the right hand side of 
the differential equation has a simple, factorized dependence 
on the regularization parameter $\epsilon$. 
While the statement has not been rigorously proved, it is expected to be true
at least for those cases that can be expressed in terms of Chen iterated integrals.
The differential equations for the canonical basis assume the following 
form 
\begin{equation}
 \frac{\partial\mathbf{I}'(\mathbf{x},\epsilon)}{\partial\mathbf{x}_i}
=\epsilon\mathcal{A}'_i(\mathbf{x})\mathbf{I}'(\mathbf{x},\epsilon),
\end{equation}
so that the iterative construction of ${\bf I}'$ as series in $\epsilon$ becomes  straightforward.
General criteria to find candidate canonical integrals are given in Ref.~\cite{Henn:2014qga} and, under certain conditions for ordinary differential equations, in Ref.~\cite{Lee:2014ioa}.  We do not use 
this last algorithm in this paper; instead, we begin by constructing canonical bases for the simplest  integrals
in the set   
and gradually move to more complex ones, as described extensively in~\cite{Gehrmann:2014bfa}. 
Since the original matrices $\mathcal{A}_i$ are relatively 
sparse,  this approach turns out to be quite practical for finding the canonical basis. 

It is convenient to choose as independent variables the center of mass energy squared $s$ and the dimensionless 
ratio $\omega = -M^2/s$.  Since the dependence of any master integral on $s$ follows uniquely from its mass 
dimension, we write master integrals as
\begin{equation}
\mathbf{I}(s,\omega) = s^{-a-2\ep}\; \mathcal{I}(\omega),
\end{equation}
where $a$ is an integer determined by the canonical mass dimension of the integral. The non-trivial information is contained in the functions  $\mathcal{I}_i(\omega)$, which are dimensionless quantities. By choosing these functions to be appropriately re-scaled versions of the master integrals found by 
\texttt{Reduze2}
\begin{equation}
  \begin{array}{cc}
    \mathcal{I}_{1}(\omega) = \epsilon^2 (\epsilon-1)(-s)^{2\epsilon} \imineq{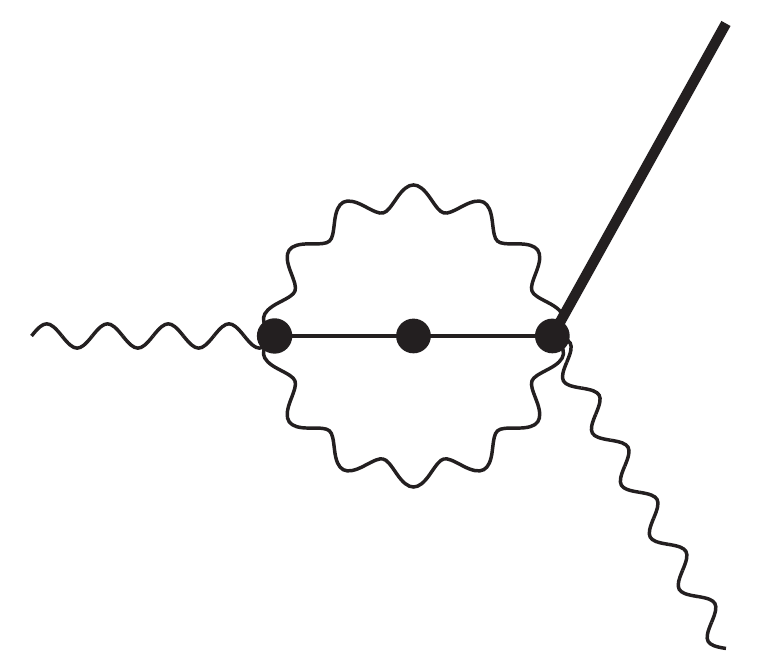}{7} &	
   \mathcal{I}_{2}(\omega) =  -\epsilon^2 (-s)^{2\epsilon+1}(\omega+1) \imineq{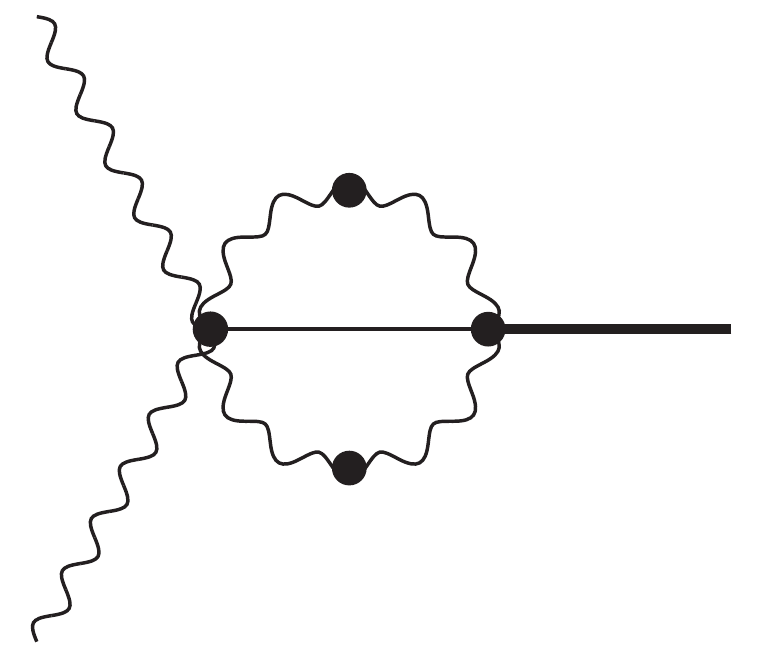}{7}	\\
   \mathcal{I}_{3}(\omega) =  -\epsilon^2 (-s)^{2\epsilon+1}(\omega+1) \imineq{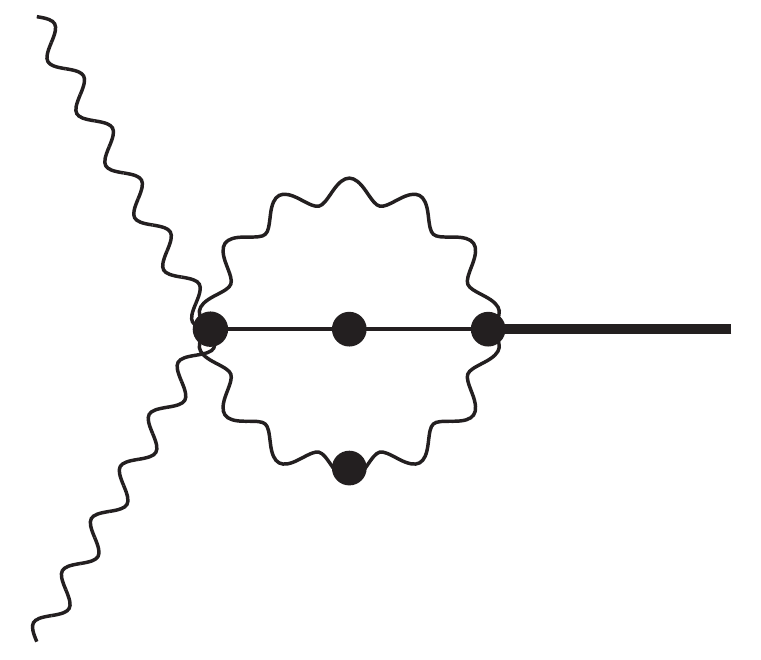}{7} &
   \mathcal{I}_{4}(\omega) =  \epsilon^3 (-s)^{2\epsilon+1} \imineq{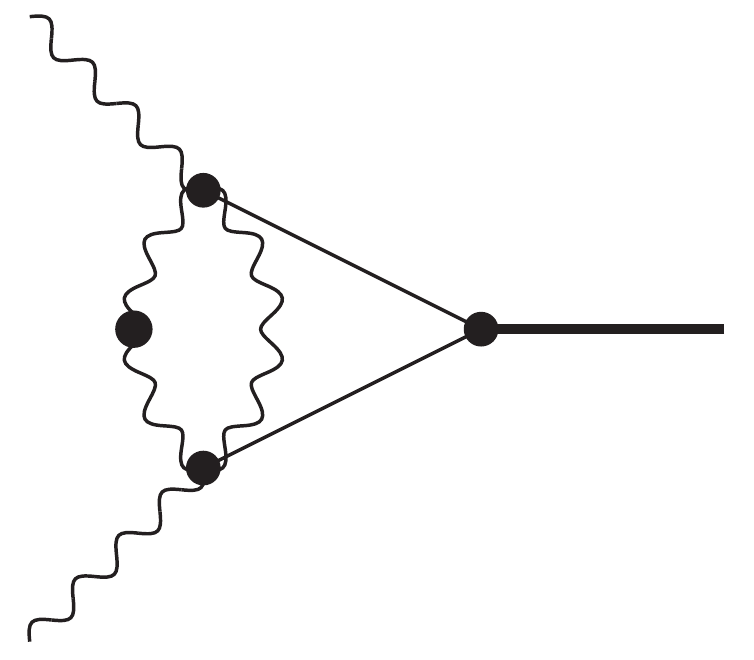}{7}	\\
   \mathcal{I}_{5}(\omega) =  \epsilon^2 (-s)^{2\epsilon+2}\omega \imineq{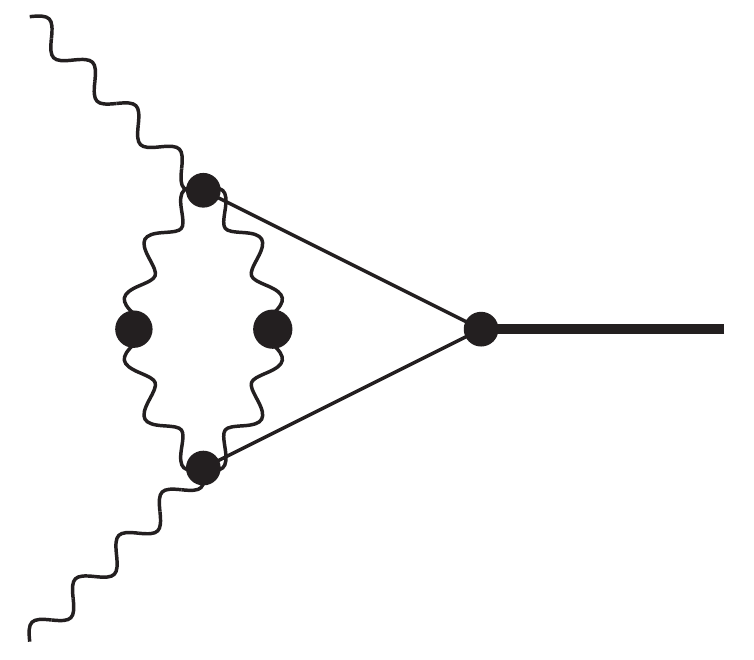}{7} & 
  \mathcal{I}_{6}(\omega) =   -\epsilon^2 (-s)^{2\epsilon+2} \imineq{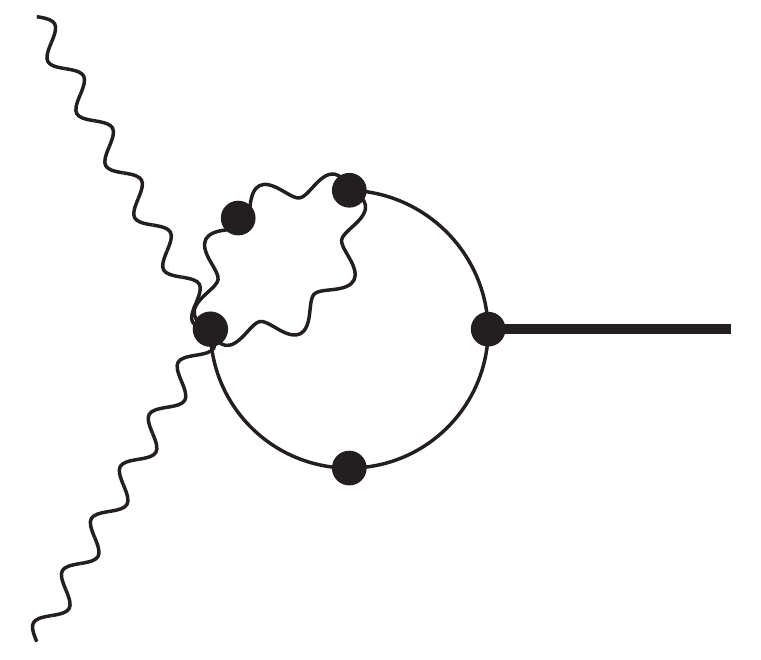}{7}	\\
   \mathcal{I}_{7}(\omega) =  \epsilon^4 (-s)^{2\epsilon+1} \imineq{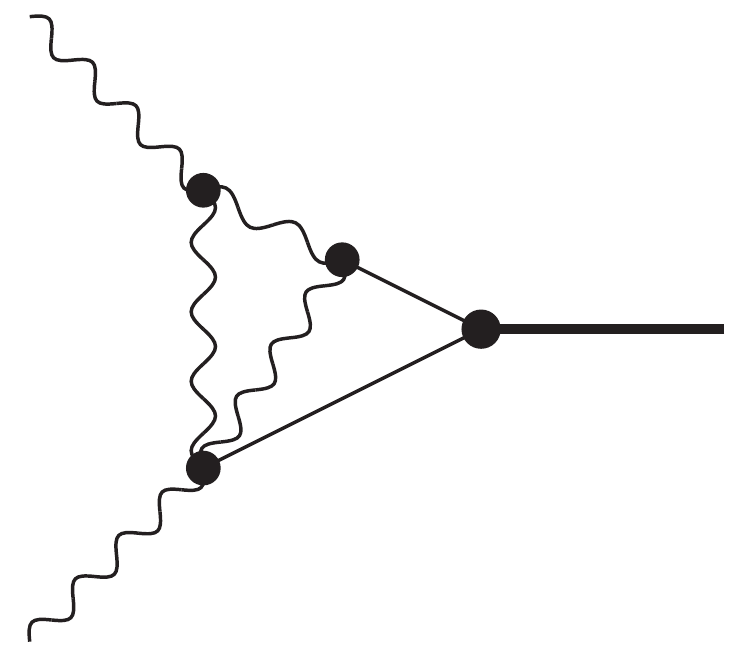}{7}	&
    \mathcal{I}_{8}(\omega) = \epsilon^4 (-s)^{2\epsilon+1} \imineq{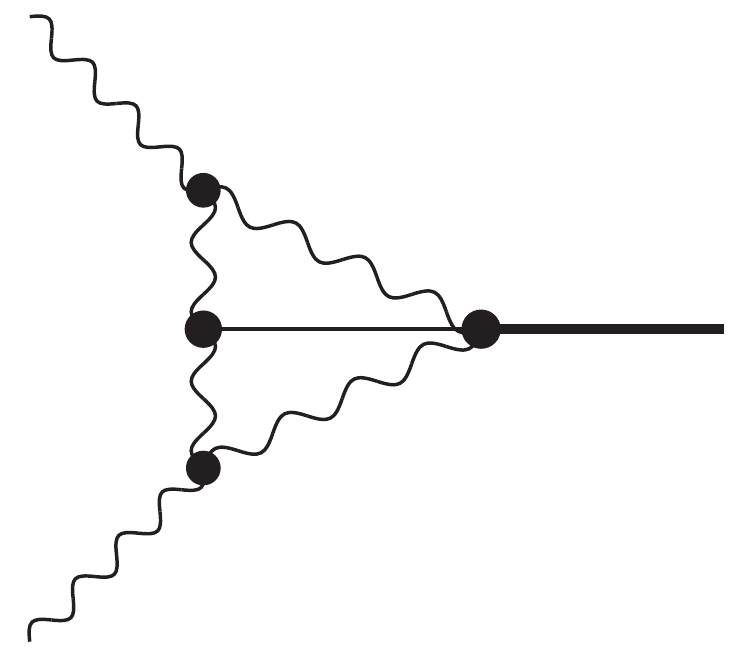}{7}	\\
   \mathcal{I}_{9}(\omega) =  \epsilon^4 (-s)^{2\epsilon+2} (4\omega+1) \imineq{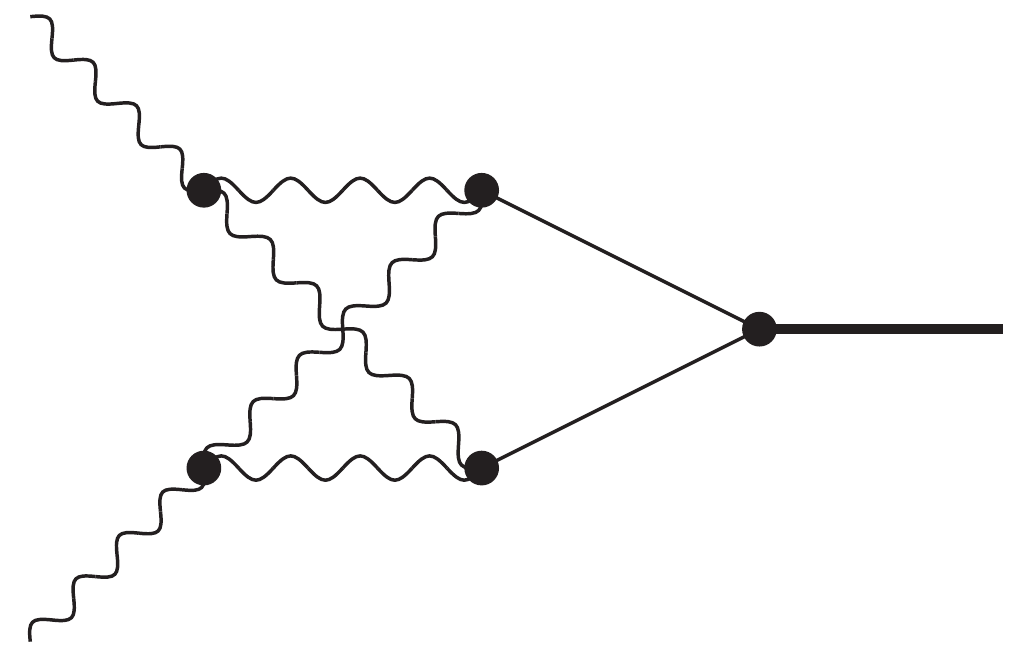}{7}	&
    \mathcal{I}_{10}(\omega) = \epsilon^2 (-s)^{2\epsilon+1} \imineq{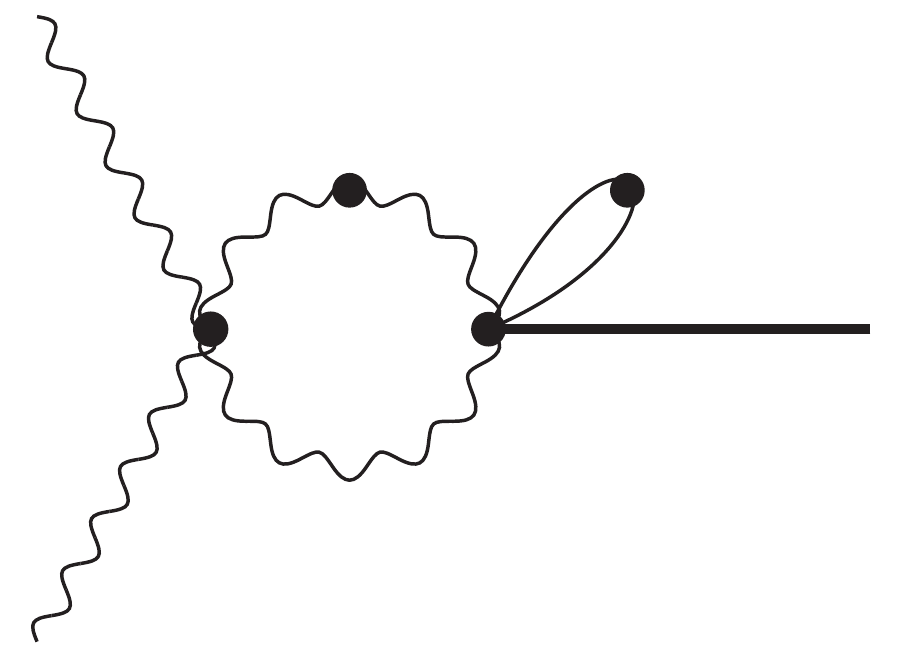}{7}	\\
  \mathcal{I}_{11}(\omega) =   -\epsilon^2 (-s)^{2\epsilon+2} \imineq{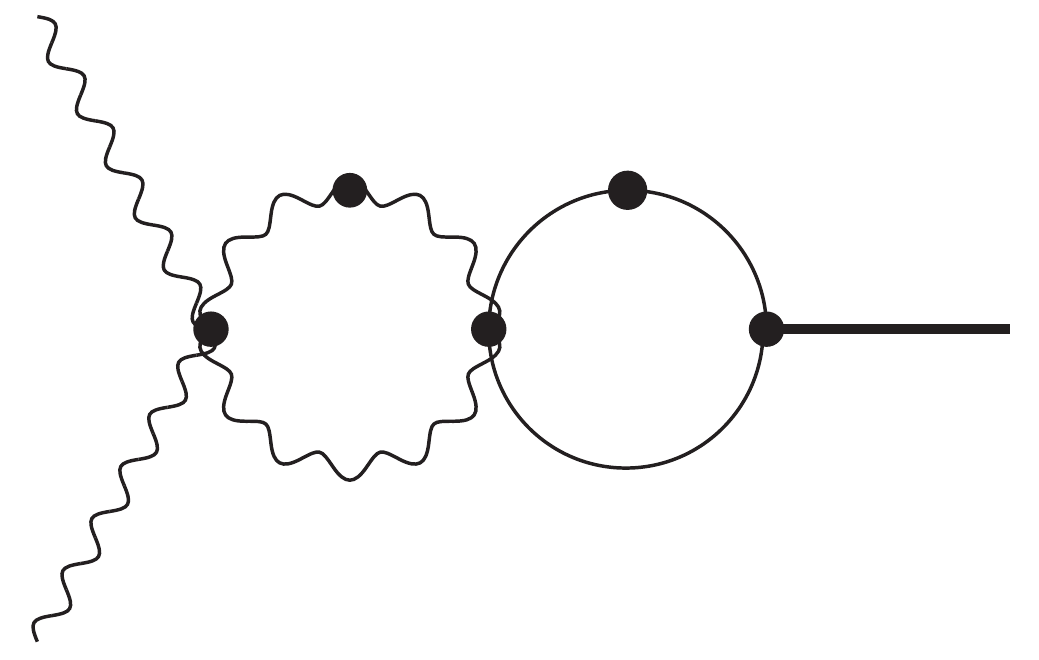}{7}	& 
   \mathcal{I}_{12}(\omega) =  -\epsilon^2 (1-2\epsilon) (-s)^{2\epsilon+2} \omega \imineq{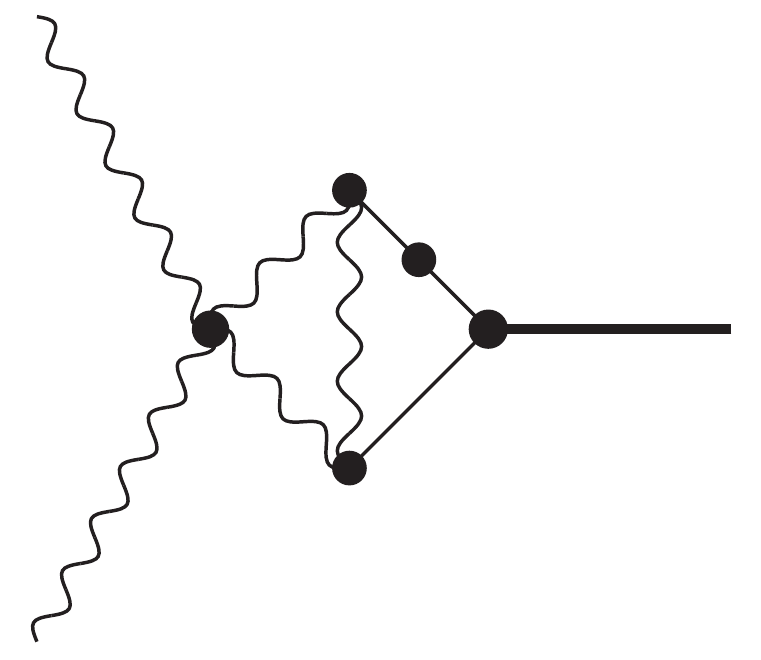}{7},
  \end{array}
\end{equation}
we can cast the system of differential equations for ${\cal I}(\omega)$ in  the following 
form 
\begin{equation}
 \frac{\mathrm{d}\mathcal{I}(\omega)}{\mathrm{d}\omega}=\left[\mathcal{A}_0(\omega)+\epsilon\mathcal{A}_1(\omega)\right]\mathcal{I}(\omega).
\end{equation}
The matrices ${\cal A}_{0,1}$ are rational functions of $\omega$, and have a block-triangular structure.

To construct a systematic expansion of master integrals in $\epsilon$, it is convenient 
to change basis of master integrals and transform the system of differential equations into a canonical 
form.  This requires  $\mathcal{A}_0$ to be removed. 
We can do that 
in a symbolic form by first solving the matrix differential equation 
\begin{equation}
 \frac{\mathrm{d} \hat{S}_{\mathcal{A}_0} }{\mathrm{d}\omega}
=\mathcal{A}_0(\omega) \hat{S}_{\mathcal{A}_0 } \;\;\;\; \rightarrow \;\;\;\;
 \hat{S}_{\mathcal{A}_0} = P_{\omega} e^{\int
 \mathcal{A}_0(\omega')\,\mathrm{d}\omega' },
\end{equation}
where $P_{\omega}$ is the path-ordering operator defined as
\begin{equation}
	P_{\omega} e^{\int A_0(\omega')\,\mathrm{d}\omega'}=\sum_{k=0}^{+\infty}
\int_{\omega_0}^{\omega}\mathcal{A}_0(\omega_1)\int_{\omega_0}^{\omega_1}\mathcal{A}_0(\omega_2)
\dots\int_{\omega_0}^{\omega_{k-1}}\mathcal{A}_0(\omega_k)\,\mathrm{d}\omega_k\dots\mathrm{d}\omega_2\mathrm{d}\omega_1.
\end{equation}
By defining a new set of master 
integrals 
\begin{equation}
\mathcal{F} = \hat{S}_{\mathcal{A}_0}^{-1}\; {\mathcal I},
\end{equation}
it is easy to see that ${\cal F}$ satisfies  differential equations  in the canonical form
\begin{equation}
 \frac{\mathrm{d}\mathcal{F}(\omega)}{\mathrm{d}\omega}=\epsilon \hat{S}_{\mathcal{A}_0}^{-1} 
\mathcal{A}_1 \hat{S}_{\mathcal{A}_0} \mathcal{F}(\omega).
\end{equation}
The non-trivial part of this procedure is to find the 
matrix $\hat S_{\mathcal{A}_0}$. 
A systematic way to do that, based on the Magnus exponentiation was suggested in Ref.~\cite{Argeri:2014qva}.
Instead, we do that iteratively, using the sparse nature of the matrix ${\cal A}_0$ and 
considering different blocks of ${\mathcal A}_{0}$ separately. 
As an illustration, consider two integrals from the list of master integrals, 
${\cal I}_2$ and ${\cal I}_3$. Neglecting the matrix ${\mathcal A}_1$, we find that they satisfy the system of 
coupled differential equations 
 \begin{equation}
 \frac{\mathrm{d}}{\mathrm{d}\omega}
 \begin{pmatrix}
  {\cal I}_2	\\
  {\cal I}_3
 \end{pmatrix}
  =
  \frac{1}{\omega+1}
  \begin{pmatrix}
0  &\,\, & -2 \\
0 & \,\, &1 \\
  \end{pmatrix}
  \begin{pmatrix}
  {\cal I}_2	\\
  {\cal I}_3
 \end{pmatrix}.
 \end{equation}
Integrating this equation, we find 
\begin{equation}
 \begin{pmatrix}
  {\cal I}_2	\\
  {\cal I}_3
 \end{pmatrix}
 =
 \hat{S}_{0}
\begin{pmatrix}
  \mathcal{C}_1	\\
  \mathcal{C}_2	\\
 \end{pmatrix},
\;\;\;
\hat{S}_{0} =  
\begin{pmatrix}
 -1 &\,\, & -2 \omega  \\
 0 &\,\, & \omega +1 \\
 \end{pmatrix}
,
\end{equation}
where $\mathcal{C}_{1,2}$ are the two integration constants. 
Since the above solution satisfies the system of differential equations for arbitrary $\mathcal{C}_1$, $\mathcal{C}_2$, the 
matrix $\hat{S}_0$ satisfies the original differential equation
\begin{equation}
 \frac{\mathrm{d} \hat{S}_{0} }{\mathrm{d}\omega}
=\frac{1}{\omega +1}   \begin{pmatrix}
0  &\,\, & -2 \\
0 &\,\, & 1 \\
  \end{pmatrix} 
 \hat{S}_{0 }
\end{equation}
and, therefore, can be taken to be a part  of $\hat{S}_{\mathcal{A}_0}$.  Finally, we 
compute $\mathcal{F} = \hat{S}_0^{-1} \mathcal{I}$ and find  
\begin{align*}
&  \mathcal{F}_2 = - \mathcal{I}_2 + \frac{2 \omega}{1+\omega} \mathcal{ I}_3 
=-\epsilon ^2 (-s)^{2\epsilon+1} \left[(\omega+1)\imineq{sbubble1}{7}+2\omega\imineq{sbubble2}{7} \right],	\\
&  \mathcal{F}_3
= 
\frac{1}{1+\omega} \mathcal{I}_3 
=\epsilon ^2 (-s)^{2\epsilon+1} \imineq{sbubble2}{7}.
\end{align*}
The system of differential equations for the integrals ${\cal F}_{2,3}$ is then 
guaranteed to be in the canonical form. We find  
\begin{equation}
 \frac{\mathrm{d}}{\mathrm{d}\omega}
 \begin{pmatrix}
  \mathcal{F}_2	\\
  \mathcal{F}_3
 \end{pmatrix}
  =
  \epsilon
  \begin{pmatrix}
 -\frac{2}{\omega +1} &\,\, & \frac{4}{\omega +1} \\
 \frac{1}{\omega +1}-\frac{1}{\omega } &\,\, & -\frac{2}{\omega +1}-\frac{1}{\omega } \\
  \end{pmatrix}
  \begin{pmatrix}
  \mathcal{F}_2	\\
  \mathcal{F}_3
 \end{pmatrix},
\end{equation}

We apply this procedure block by block, 
to the block-triangular matrix $\mathcal{A}_0+\epsilon\mathcal{A}_1$ and 
obtain the  canonical system of differential equations that 
we write in the following  form 
\begin{equation}
 \mathrm{d}\mathcal{F}(\omega)  = \epsilon \,{\rm d} \mathcal{B}(\omega)\mathcal{F}(\omega).
\label{eq3.17}
\end{equation}
The canonical basis of the master integrals $\mathcal{F}(\omega)$ reads 
\begin{bigeq}
  \begin{tiny}
  \begin{pmatrix}
    \epsilon^2 (\epsilon-1)(-s)^{2\epsilon} \imineq{0bubble}{7}	\\
    
    -\epsilon ^2 (-s)^{2\epsilon+1} \left[(\omega+1)\imineq{sbubble1}{7}+2\omega\imineq{sbubble2}{7} \right] \\
    
    \epsilon ^2 (-s)^{2\epsilon+1} \imineq{sbubble2}{7}	\\
    
    \epsilon^3 (-s)^{2\epsilon+1} \imineq{Triangle1}{7}	\\
    
    \epsilon^2 \left[(1-\epsilon)\frac{2}{\sqrt{1+4\omega}} (-s)^{2\epsilon}\imineq{0bubble}{7}+\epsilon\frac{3(1+2\omega)}{\sqrt{1+4\omega}}(-s)^{2\epsilon+1} \imineq{Triangle1}{7}-(-s)^{2\epsilon+2}\frac{\omega^2}{\sqrt{1+4\omega}}\imineq{Triangle2}{7}\right]	\\
    
    \epsilon^2\left[(1-\epsilon)(-s)^{2\epsilon}\frac{\sqrt{1+4\omega}}{2} \imineq{0bubble}{7}+(-s)^{2\epsilon+1} \frac{(\omega+1)\sqrt{1+4\omega}}{4}\imineq{sbubble1}{7}+(-s)^{2\epsilon+1} \frac{(\omega+1)\sqrt{1+4\omega}}{2} \imineq{sbubble2}{7}+(-s)^{2\epsilon+2} \omega\sqrt{1+4\omega}\imineq{Inclusion}{7}\right]	\\
    
    \epsilon^4 (-s)^{2\epsilon+1} \imineq{DoubleTriangle}{7} \\
    
    \epsilon^4 (-s)^{2\epsilon+1} \imineq{Arrow}{7}  \\
    
    \epsilon^4 (-s)^{2\epsilon+2} \sqrt{4\omega+1} \imineq{NonPlanar}{7}   \\
    
    \epsilon^2 (-s)^{2\epsilon+1} \imineq{btadpole}{7}  \\
    
    -\epsilon^2 (-s)^{2\epsilon+2} \sqrt{4\omega+1} \imineq{doublebubble}{7}   \\
    
    \epsilon^2\left[\frac{1-\epsilon}{2}(-s)^{2\epsilon} \imineq{0bubble}{7}+ (-s)^{2\epsilon+1} \frac{\omega+1}{4} \imineq{sbubble1}{7}+(-s)^{2\epsilon+1} \frac{\omega+1}{2} \imineq{sbubble2}{7}+(-s)^{2\epsilon+2} \omega \imineq{Inclusion}{7}+ (-s)^{2\epsilon+2} \imineq{doublebubble}{7}+(1-2\epsilon) (-s)^{2\epsilon+2} \omega \imineq{Kite}{7}\right]	\\
  \end{pmatrix}.
  \end{tiny}
\end{bigeq}
For  the matrix  $\mathcal{B}(\omega)$  we obtain \footnote{The 
signs of the arguments of the logarithms are chosen to ensure that for positive $\omega$ the 
logarithms are real.} 
\begin{multline}
 \mathcal{B}(\omega)=B_1\,\log\omega+
		    B_2\, \log(1+\omega)+	\\
		    +B_3\,[ \log(-1+\sqrt{1+4\omega})- \log(1+\sqrt{1+4\omega})]
                 +B_4\,\log(1+4\omega).
\label{eq8}
\end{multline}
The  $\omega$-independent matrix coefficients in the above equation are given by  
\begin{align}
 &\tiny B_1=
 \begin{pmatrix}
 -2 & 0 & 0 & 0 & 0 & 0 & 0 & 0 & 0 & 0 & 0 & 0 \\
 0 & 0 & 0 & 0 & 0 & 0 & 0 & 0 & 0 & 0 & 0 & 0 \\
 0 & -1 & -1 & 0 & 0 & 0 & 0 & 0 & 0 & 0 & 0 & 0 \\
 0 & 0 & 0 & -3 & 0 & 0 & 0 & 0 & 0 & 0 & 0 & 0 \\
 0 & 0 & 0 & 0 & -1 & 0 & 0 & 0 & 0 & 0 & 0 & 0 \\
 0 & 0 & 0 & 0 & 0 & -1 & 0 & 0 & 0 & 0 & 0 & 0 \\
 \frac{1}{2} & \frac{1}{4} & \frac{1}{2} & -\frac{1}{2} & 0 & 0 & -2 & 0 & 0 & 0 & 0 & 0 \\
 -1 & -\frac{1}{2} & 0 & 0 & 0 & 0 & 0 & -2 & 0 & 0 & 0 & 0 \\
 0 & 0 & 0 & 0 & 0 & 0 & 0 & 0 & 0 & 0 & 0 & 0 \\
 0 & 0 & 0 & 0 & 0 & 0 & 0 & 0 & 0 & -1 & 0 & 0 \\
 0 & 0 & 0 & 0 & 0 & 0 & 0 & 0 & 0 & 0 & 0 & 0 \\
 1 & \frac{1}{2} & \frac{1}{2} & 0 & 0 & 0 & 0 & 0 & 0 & 0 & 0 & -2 \\
 \end{pmatrix},
 \quad
 \tiny B_2=
 \begin{pmatrix}
 0 & 0 & 0 & 0 & 0 & 0 & 0 & 0 & 0 & 0 & 0 & 0 \\
 0 & -2 & 4 & 0 & 0 & 0 & 0 & 0 & 0 & 0 & 0 & 0 \\
 0 & 1 & -2 & 0 & 0 & 0 & 0 & 0 & 0 & 0 & 0 & 0 \\
 0 & 0 & 0 & 0 & 0 & 0 & 0 & 0 & 0 & 0 & 0 & 0 \\
 0 & 0 & 0 & 0 & 0 & 0 & 0 & 0 & 0 & 0 & 0 & 0 \\
 0 & 0 & 0 & 0 & 0 & 0 & 0 & 0 & 0 & 0 & 0 & 0 \\
 0 & 0 & 0 & 0 & 0 & 0 & 0 & 0 & 0 & 0 & 0 & 0 \\
 0 & 0 & 0 & 0 & 0 & 0 & 0 & 0 & 0 & 0 & 0 & 0 \\
 0 & 0 & 0 & 0 & 0 & 0 & 0 & 0 & 0 & 0 & 0 & 0 \\
 0 & 0 & 0 & 0 & 0 & 0 & 0 & 0 & 0 & 0 & 0 & 0 \\
 0 & 0 & 0 & 0 & 0 & 0 & 0 & 0 & 0 & 0 & 0 & 0 \\
 0 & 0 & 0 & 0 & 0 & 0 & 0 & 0 & 0 & 0 & 0 & 0 \\
 \end{pmatrix},
\\\noindent&\tiny B_3=
 \begin{pmatrix}
 0 & 0 & 0 & 0 & 0 & 0 & 0 & 0 & 0 & 0 & 0 & 0 \\
 0 & 0 & 0 & 0 & 0 & 0 & 0 & 0 & 0 & 0 & 0 & 0 \\
 0 & 0 & 0 & 0 & 0 & 0 & 0 & 0 & 0 & 0 & 0 & 0 \\
 0 & 0 & 0 & 0 & 1 & 0 & 0 & 0 & 0 & 0 & 0 & 0 \\
 -4 & 0 & 0 & 3 & 0 & 0 & 0 & 0 & 0 & 0 & 0 & 0 \\
 \frac{1}{2} & -\frac{3}{4} & 0 & 0 & 0 & 0 & 0 & 0 & 0 & 0 & 0 & 0 \\
 0 & 0 & 0 & 0 & -\frac{1}{2} & -1 & 0 & 0 & 0 & 0 & 0 & 0 \\
 0 & 0 & 0 & 0 & 0 & 0 & 0 & 0 & 0 & 0 & 0 & 0 \\
 6 & 3 & 0 & -8 & 0 & 0 & -8 & 4 & 0 & 0 & 0 & 0 \\
 0 & 0 & 0 & 0 & 0 & 0 & 0 & 0 & 0 & 0 & 0 & 0 \\
 0 & 0 & 0 & 0 & 0 & 0 & 0 & 0 & 0 & -1 & 0 & 0 \\
 0 & 0 & 0 & 0 & 0 & 2 & 0 & 0 & 0 & 0 & -2 & 0 \\
 \end{pmatrix},
 \quad
 \tiny B_4=
 \begin{pmatrix}
 0 & 0 & 0 & 0 & 0 & 0 & 0 & 0 & 0 & 0 & 0 & 0 \\
 0 & 0 & 0 & 0 & 0 & 0 & 0 & 0 & 0 & 0 & 0 & 0 \\
 0 & 0 & 0 & 0 & 0 & 0 & 0 & 0 & 0 & 0 & 0 & 0 \\
 0 & 0 & 0 & 0 & 0 & 0 & 0 & 0 & 0 & 0 & 0 & 0 \\
 0 & 0 & 0 & 0 & -1 & 0 & 0 & 0 & 0 & 0 & 0 & 0 \\
 0 & 0 & 0 & 0 & 0 & -1 & 0 & 0 & 0 & 0 & 0 & 0 \\
 0 & 0 & 0 & 0 & 0 & 0 & 0 & 0 & 0 & 0 & 0 & 0 \\
 0 & 0 & 0 & 0 & 0 & 0 & 0 & 0 & 0 & 0 & 0 & 0 \\
 0 & 0 & 0 & 0 & 0 & 0 & 0 & 0 & -2 & 0 & 0 & 0 \\
 0 & 0 & 0 & 0 & 0 & 0 & 0 & 0 & 0 & 0 & 0 & 0 \\
 0 & 0 & 0 & 0 & 0 & 0 & 0 & 0 & 0 & 0 & -1 & 0 \\
 0 & 0 & 0 & 0 & 0 & 0 & 0 & 0 & 0 & 0 & 0 & 0 \\
 \end{pmatrix}.
\end{align}

It is convenient to remove the  square roots from the Eqs.~(\ref{eq3.17},\ref{eq8}) by changing variables  $\omega \to y$ where\footnote{The relations between $s$, $\omega$ and $y$ are summarized in Table \ref{tabvar}.}
\begin{equation}
 y =\frac{\sqrt{1+4\omega}-1}{\sqrt{1+4\omega}+1}.
\end{equation}
The differential equations~(\ref{eq3.17}) take the following form 
\begin{equation}
 \mathrm{d}\mathcal{F}(y)  = \epsilon\, {\rm d} \mathcal{C}(y)\mathcal{F}(y),
 \label{eq3.18}
\end{equation}
where  the matrix $\mathcal{C}$ reads 
\begin{equation}
\mathcal{C}(y)=C_1\, \log y+
		    C_2\, \log(1-y)+
		    C_3\, \log(1+y)+
		    C_4\, \log(1-y+y^2).
\label{eq3.22}
\end{equation}
The $y$-independent matrices $C_{1, \dots ,4}$ are
\begin{align}
&\tiny
 C_1=
 \begin{pmatrix}
 -2 & 0 & 0 & 0 & 0 & 0 & 0 & 0 & 0 & 0 & 0 & 0 \\
 0 & 0 & 0 & 0 & 0 & 0 & 0 & 0 & 0 & 0 & 0 & 0 \\
 0 & -1 & -1 & 0 & 0 & 0 & 0 & 0 & 0 & 0 & 0 & 0 \\
 0 & 0 & 0 & -3 & 1 & 0 & 0 & 0 & 0 & 0 & 0 & 0 \\
 -4 & 0 & 0 & 3 & -1 & 0 & 0 & 0 & 0 & 0 & 0 & 0 \\
 \frac{1}{2} & -\frac{3}{4} & 0 & 0 & 0 & -1 & 0 & 0 & 0 & 0 & 0 & 0 \\
 \frac{1}{2} & \frac{1}{4} & \frac{1}{2} & -\frac{1}{2} & -\frac{1}{2} & -1 & -2 & 0 & 0 & 0 & 0 & 0 \\
 -1 & -\frac{1}{2} & 0 & 0 & 0 & 0 & 0 & -2 & 0 & 0 & 0 & 0 \\
 6 & 3 & 0 & -8 & 0 & 0 & -8 & 4 & 0 & 0 & 0 & 0 \\
 0 & 0 & 0 & 0 & 0 & 0 & 0 & 0 & 0 & -1 & 0 & 0 \\
 0 & 0 & 0 & 0 & 0 & 0 & 0 & 0 & 0 & -1 & 0 & 0 \\
 1 & \frac{1}{2} & \frac{1}{2} & 0 & 0 & 2 & 0 & 0 & 0 & 0 & -2 & -2 \\
 \end{pmatrix},
 \quad
 \tiny C_2=
 \begin{pmatrix}
 4 & 0 & 0 & 0 & 0 & 0 & 0 & 0 & 0 & 0 & 0 & 0 \\
 0 & 4 & -8 & 0 & 0 & 0 & 0 & 0 & 0 & 0 & 0 & 0 \\
 0 & 0 & 6 & 0 & 0 & 0 & 0 & 0 & 0 & 0 & 0 & 0 \\
 0 & 0 & 0 & 6 & 0 & 0 & 0 & 0 & 0 & 0 & 0 & 0 \\
 0 & 0 & 0 & 0 & 4 & 0 & 0 & 0 & 0 & 0 & 0 & 0 \\
 0 & 0 & 0 & 0 & 0 & 4 & 0 & 0 & 0 & 0 & 0 & 0 \\
 -1 & -\frac{1}{2} & -1 & 1 & 0 & 0 & 4 & 0 & 0 & 0 & 0 & 0 \\
 2 & 1 & 0 & 0 & 0 & 0 & 0 & 4 & 0 & 0 & 0 & 0 \\
 0 & 0 & 0 & 0 & 0 & 0 & 0 & 0 & 4 & 0 & 0 & 0 \\
 0 & 0 & 0 & 0 & 0 & 0 & 0 & 0 & 0 & 2 & 0 & 0 \\
 0 & 0 & 0 & 0 & 0 & 0 & 0 & 0 & 0 & 0 & 2 & 0 \\
 -2 & -1 & -1 & 0 & 0 & 0 & 0 & 0 & 0 & 0 & 0 & 4 \\
 \end{pmatrix},
 \\\noindent&\tiny
 C_3=
 \begin{pmatrix}
 0 & 0 & 0 & 0 & 0 & 0 & 0 & 0 & 0 & 0 & 0 & 0 \\
 0 & 0 & 0 & 0 & 0 & 0 & 0 & 0 & 0 & 0 & 0 & 0 \\
 0 & 0 & 0 & 0 & 0 & 0 & 0 & 0 & 0 & 0 & 0 & 0 \\
 0 & 0 & 0 & 0 & 0 & 0 & 0 & 0 & 0 & 0 & 0 & 0 \\
 0 & 0 & 0 & 0 & -2 & 0 & 0 & 0 & 0 & 0 & 0 & 0 \\
 0 & 0 & 0 & 0 & 0 & -2 & 0 & 0 & 0 & 0 & 0 & 0 \\
 0 & 0 & 0 & 0 & 0 & 0 & 0 & 0 & 0 & 0 & 0 & 0 \\
 0 & 0 & 0 & 0 & 0 & 0 & 0 & 0 & 0 & 0 & 0 & 0 \\
 0 & 0 & 0 & 0 & 0 & 0 & 0 & 0 & -4 & 0 & 0 & 0 \\
 0 & 0 & 0 & 0 & 0 & 0 & 0 & 0 & 0 & 0 & 0 & 0 \\
 0 & 0 & 0 & 0 & 0 & 0 & 0 & 0 & 0 & 0 & -2 & 0 \\
 0 & 0 & 0 & 0 & 0 & 0 & 0 & 0 & 0 & 0 & 0 & 0 \\
 \end{pmatrix},
  \quad
 \tiny C_4=
 \begin{pmatrix}
 0 & 0 & 0 & 0 & 0 & 0 & 0 & 0 & 0 & 0 & 0 & 0 \\
 0 & -2 & 4 & 0 & 0 & 0 & 0 & 0 & 0 & 0 & 0 & 0 \\
 0 & 1 & -2 & 0 & 0 & 0 & 0 & 0 & 0 & 0 & 0 & 0 \\
 0 & 0 & 0 & 0 & 0 & 0 & 0 & 0 & 0 & 0 & 0 & 0 \\
 0 & 0 & 0 & 0 & 0 & 0 & 0 & 0 & 0 & 0 & 0 & 0 \\
 0 & 0 & 0 & 0 & 0 & 0 & 0 & 0 & 0 & 0 & 0 & 0 \\
 0 & 0 & 0 & 0 & 0 & 0 & 0 & 0 & 0 & 0 & 0 & 0 \\
 0 & 0 & 0 & 0 & 0 & 0 & 0 & 0 & 0 & 0 & 0 & 0 \\
 0 & 0 & 0 & 0 & 0 & 0 & 0 & 0 & 0 & 0 & 0 & 0 \\
 0 & 0 & 0 & 0 & 0 & 0 & 0 & 0 & 0 & 0 & 0 & 0 \\
 0 & 0 & 0 & 0 & 0 & 0 & 0 & 0 & 0 & 0 & 0 & 0 \\
 0 & 0 & 0 & 0 & 0 & 0 & 0 & 0 & 0 & 0 & 0 & 0 \\
 \end{pmatrix}.
\end{align}

It is straightforward to write the solution 
of the system of differential equations~(\ref{eq3.18})
as Taylor series in $\epsilon$
\begin{equation}
  \label{DY}
  \begin{split}
 \mathcal{F}(y)	&=\mathcal{F}_0^{(0)}
+\epsilon\left[\int_0^y\mathcal{C}(\tau_1)\mathcal{F}_0^{(0)}\,\mathrm{d}\tau_1+\mathcal{F}_0^{(1)}\right]+	\\
		&+\epsilon^2\left[\int_0^y\mathcal{C}(\tau_1)\int_0^{\tau_1}\mathcal{C}(\tau_2)\mathcal{F}_0^{(0)}\,\mathrm{d}\tau_2\mathrm{d}\tau_1+\int_0^y\mathcal{C}(\tau_1)\mathcal{F}_0^{(1)}\,\mathrm{d}\tau_1+\mathcal{F}_0^{(2)}\right]+...,
  \end{split}
\end{equation}
where $\mathcal{F}_0^{(i)}$ are integration constants that can not be fixed 
from the differential equations.

Given the iterative structure of the solution, it can be written as a linear combination of 
the so-called 
Goncharov's  polylogarithms (GPLs), also known as hyperlogarithms,
defined as~\cite{Goncharov, Remiddi:1999ew, Gehrmann:2000zt, Vollinga:2004sn} 
\begin{equation}
 G(m_w,\mathbf{m}_{w-1};x):=
 \begin{cases}
  \frac{1}{w!}\log^wx	&\text{if $\mathbf{m}_w=(0,\dots,0)$}	\\
  \int_0^x f(m_w;\tau) \; G(\mathbf{m}_{w-1};\tau) \,\mathrm{d}\tau	
&\text{if $\mathbf{m}_w\neq(0,\dots,0)$}
 \end{cases},
\end{equation}
where $\mathbf{m}_w$ indicates the vector $(m_w,\mathbf{m}_{w-1})$. The functions $f(a;\tau)$ represent the \emph{integration kernels}; for our system of 
differential equations they span the following set 
\begin{equation}
 f(0;\tau)=\frac{1}{\tau}, \quad
 f(1;\tau)=\frac{1}{\tau-1}, \quad
 f(-1;\tau)=\frac{1}{\tau+1}, \quad
 f(r;\tau)=\frac{2\tau-1}{\tau^2-\tau+1}.
\end{equation}
The last term in the set is quadratic in the integration variable; it is possible to re-write 
it in a usual  linear form 
\begin{equation}
 f(r;\tau)=\frac{2\tau-1}{\tau^2-\tau+1}=\frac{1}{\tau - r_+} + \frac{1}{\tau - r_-}=f(r_+;\tau)+f(r_-;\tau), 
\;\;\;\; r_{\pm} = \mathrm{e}^{\pm \mathrm{i}\frac{\mathrm{\pi}}{3}},
\end{equation}  
at the expense of introducing complex-valued poles. This last step is essential for numerical 
evaluation of the GPLs but it is not required to integrate  the 
system of differential equations  since, thanks to the linearity of 
the differential equations and the definition  of the GPLs,
\begin{equation}
 G(\dots,r,\dots;x)=G(\dots,r_-,\dots;x)+G(\dots,r_+,\dots;x).
\label{eq3.30}
\end{equation}
This implies that we can  perform the analytic integration using the  symbol $r$ and then, for the numerical 
evaluation of the final result, switch to  $r_{\pm}$ using 
Eq.~(\ref{eq3.30}).\footnote{The generalization of GPLs studied here has been already 
considered in detail in~\cite{Ablinger:2011te, vonManteuffel:2013vja}.}

~

In Table~\ref{tabvar} the relations among $s$, $\omega$ and $y$, as well as the different kinematic regions, are summarized.

\begin{table}[H]
\centering
\begin{tabular}{cccc}
\toprule
  Variable						&Euclidean		&Minkowski, below threshold					&Above threshold	\\
\midrule
  $s$							&$-\infty<s<0$		&$0<s<4M^2$							&$4M^2<s<+\infty$		\\
  $\omega=-\frac{M^2}{s}$				&$0<\omega<+\infty$	&$-\infty<\omega<-\frac{1}{4}$					&$-\frac{1}{4}<\omega<0$	\\
  $y=\frac{\sqrt{1+4\omega}-1}{\sqrt{1+4\omega}+1}$	&$0<y<1$		&$\mathrm{e}^{\mathrm{i}\theta}$, $0<\theta<\mathrm{\pi}$	&$-1<y<0$			\\
\bottomrule
\end{tabular}
\caption{Different kinematic regions in $s$, $\omega$ and $y$.}
\label{tabvar}
\end{table}

\section{Boundary conditions and analytic continuation}
\label{BC}\setcounter{equation}{0} 
\numberwithin{equation}{section}

Linear differential equations allow us to restore the dependencies 
of master integrals on $\omega$ up to a single constant. The constant 
can  be fixed by computing the required integral at any  point $\omega = \omega_0$ 
and comparing the result with the solution of the differential equation. 

It turns out to be convenient 
to determine the boundary conditions by 
computing the master integrals in    $\omega  \to \infty$ or $y \to 1$ limit. 
Since $\omega = -M^2/s$, this corresponds to 
$M^2 \to \infty$ at fixed $s$;  in  this  limit  the integrals 
can be easily computed using the so-called large-mass 
expansion procedure \cite{Smirnov:2002pj}. 
The large-mass expansion procedure can be formulated as follows: 
consider two different scalings for each of the 
loop momenta $k_{1,2} \sim M$ and $k_{1,2} \sim \sqrt{s}$ 
and, for a chosen  scaling, systematically 
expand the 
integrand in Taylor series in all small variables.  The set of small 
variables will differ from scaling to scaling; for example, if 
$k_{1} \sim k_{2} \sim M$, one Taylor expands in external momenta 
but if the scaling $k_{1} \sim M$, $k_{2} \sim \sqrt{s}$ is considered, 
one Taylor expands in the external momenta and $k_2$.
It is easy to see that, 
to leading order in $s/M^2$,  most of the master integrals are  
expressed in terms of two-loop tadpole integrals and, 
in some cases, in terms of  products of one-loop three- and two-point 
integrals and one-loop tadpole integrals.

As an illustration, consider the non-planar master integral 
\begin{equation}
{\cal F}_9(\omega) =     \epsilon^4 (-s)^{2\epsilon+2} \sqrt{4\omega+1} \imineq{NonPlanar}{7}. 
\end{equation}
We are interested in determining its behavior in the $y \to 1$ limit. By applying the 
large mass expansion, we find that the non-planar integral scales as 
\begin{equation}
 \lim_{M\gg\sqrt{s}}\imineq{NonPlanar}{7}=\imineq{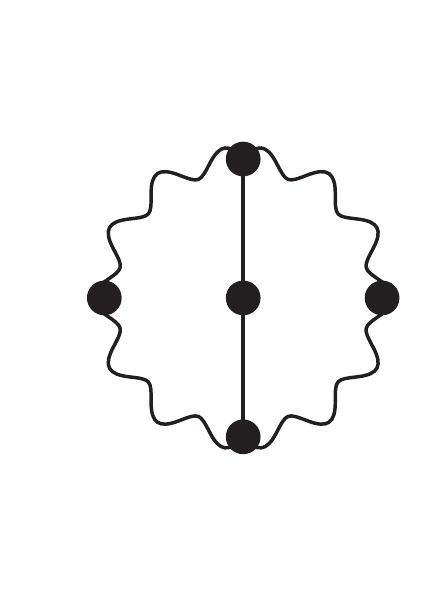}{7}={\cal O} \left ( M^{-4}  \right ) 
\end{equation} 
in the large-$M$ limit. This implies that the large-$M$ limit of the $\mathcal{F}_9$ non-planar master integral reads 
\begin{equation}
{\cal F}_9(\omega)  \sim    \frac{ \epsilon^4 (-s)^{2\epsilon+2} \sqrt{4\omega+1} }{M^4}
\sim \frac{1}{\omega^{3/2}}
\sim (1-y)^3,
\end{equation}
where we used relations among $s$, $\omega$ and $y$ variables summarized in Table~\ref{tabvar}. $\mathcal{F}_9$ vanishes as $(1-y)^3$ as $y$ goes to $1$, and this information is sufficient to fix the 
constant of integration for this master integral. 

A similar analysis reveals that only three master integrals ${\cal F}_{1}, 
{\cal F}_{2},  {\cal F}_{10}$  possess  non-vanishing $y \to 1$ limit.  These limits are 
\begin{align}
\lim_{y \to 1}\mathcal{F}_1(y)&=-(1-y)^{4\epsilon} \frac{ \Gamma(1 + 2\epsilon)\Gamma(1-\epsilon) }{ 2 \Gamma(\epsilon+1)},	\\
\lim_{y \to 1}\mathcal{F}_2(y)&=(1-y)^{4\epsilon} \frac{ \Gamma(1 + 2\epsilon)\Gamma(1-\epsilon) }{ \Gamma(\epsilon+1)},	\\
\lim_{y \to 1}\mathcal{F}_{10}(y)&=(1-y)^{4\epsilon} \frac{\Gamma^2(1-\epsilon)}{\Gamma(1-2\epsilon)}.
\end{align}

To fix the boundary conditions, we need to evaluate the GPLs of the form 
$G(\mathbf{m}, y)$, where $\mathbf{m}$ is composed of the elements of the set 
\begin{equation}
 \{ \quad 0, \quad 1, \quad -1, \quad \mathrm{e}^{\mathrm{i}\frac{\mathrm{\pi}}{3}}, \quad \mathrm{e}^{-\mathrm{i}\frac{\mathrm{\pi}}{3}} \quad \},
\label{eq4.5}
\end{equation}
and the limit $y \to 1$ is taken where possible. 
For weights higher than three,
 not all the Goncharov polylogarithms at $y = 1$ with the arguments 
from Eq.~(\ref{eq4.5}) 
can be analytically expressed in terms 
of canonical irrational 
numbers such as $\pi$ and $\zeta(n)$. Nevertheless, we expect that 
the boundary constants 
are linear combinations of  these irrational numbers; to find them 
we follow a numerical approach. We use our solution of the 
differential equations and the boundary 
conditions discussed above to find the integration constants 
numerically with high precision.\footnote{
For the evaluation of the GPLs, the GINAC implementation 
was used, see Ref.~\cite{Vollinga:2004sn}.} We then fit the resulting
numerical value to a linear combination of $\pi^2$ and $\zeta(n)$ 
of a well-defined weight. 
For example, at weight two we only have $\pi^2$, at weight 
three  $\zeta(3)$, at weight 
four  $\pi^4$, at weight five  $\zeta(5) $ and $ \pi^2 \zeta(3)$ 
and at weight six   $\pi^6$ and 
$\zeta(3)^2$.  For each of the master integrals, 
we have achieved the  matching of the numerical and the analytic 
results to at least  750 digits.

Our final remark concerns the analytic continuation 
of the master integrals $\mathcal{F}(y)$. So far, we have 
studied them in the Euclidean region but we need them in the 
region where $s = m_H^2 > 0$ and, yet, $s < 4 M^2$. 
The correct analytic continuation is achieved by replacing 
$s \to s + \mathrm{i}0$ at fixed $M^2$.  It is easy to see that 
this implies $\omega \to \omega + \mathrm{i}0$ and $ y \to y + \mathrm{i} 0$. 

~

All the master integrals evaluated here have been compared for at least four 
different values of $s/M^2$, both in the Euclidean and Minkowski region, 
to the numerical results obtained with the program 
\texttt{Secdec} \cite{Borowka:2015mxa}. In all cases we found  
excellent agreement.

\section{Form factor for $gg \to H$}
\label{ampli}\setcounter{equation}{0} 
\numberwithin{equation}{section}

The $gg \to H$ amplitude is described by a single form factor, as explained in Section~\ref{tensdec}. This form factor receives contributions from loops with $W$ and $Z$ bosons. 
The form factor is finite in four dimensions ($\epsilon \to 0$) and can be written as:
\begin{multline}
 \label{FF}
 \mathcal{F}(s,m_W^2,m_Z^2)=-(4\mathrm{\pi})^{4\epsilon} (-s)^{2\epsilon} \Gamma^2(\epsilon+1)\\
 \times\frac{\mathrm{i}\alpha_S\alpha^2}{4\mathrm{\pi}\sin^4\theta_W}\frac{v}{2}
 \left[4\mathcal{A}(y_W)
 +\frac{2}{\cos^4\theta_W}\left(\frac{5}{4}-\frac{7}{3}\sin^2\theta_W+\frac{22}{9}\sin^4\theta_W\right)\mathcal{A}(y_Z)\right],
\end{multline}
where
\begin{equation}
 y_W=\frac{\sqrt{1-4m_W^2/m_H^2}-1}{\sqrt{1-4m_W^2/m_H^2}+1}, \qquad y_Z=\frac{\sqrt{1-4m_Z^2/m_H^2}-1}{\sqrt{1-4m_Z^2/m_H^2}+1}.
\end{equation}

We take the CKM matrix to be an identity matrix. The contributions of $W$ bosons is computed in Eq.~(\ref{FF}) taking into account first and second generations. The contribution of the $Z$ boson is calculated for five massless quarks ($u$, $d$, $s$, $c$ and $b$).

The function $\mathcal{A}$ in Eq.~(\ref{FF}) can be expanded in $\epsilon$; we have computed it through $\mathcal{O}\left(\epsilon^2\right)$:
\begin{equation}
 \mathcal{A}(y)=\mathcal{A}_0(y)+\epsilon\mathcal{A}_1(y)+\epsilon^2\mathcal{A}_2(y)+\mathcal{O}
\left(\epsilon^3\right).
\end{equation}

The function $\mathcal{A}_0(y)$ reads
\begin{equation}
 \begin{split}
 &\mathcal{A}_0(y)=\frac{1}{6(y-1)^3}
 \left[-6 -6y (y^2-y+2) G(0,0,r,y)-6 (1-y) (y^2-y+1) G(r,y)	\right.	\\
 &\left.+(y+1) (y^2+1) [18 G(-1,0,r,y)+\mathrm{\pi} ^2 G(-1,y)-18 G(-1,0,0,y)]	\right.	\\
 &\left.+12y(2 y^2+y+1) [G(0,1,0,y)-G(0,1,r,y)]	\right.	\\
 &\left.+2(1-y) (y^2+y+1)[6 G(1,0,r,y)-12 G(1,1,r,y)- \mathrm{\pi} ^2 G(1,y)+12 G(1,1,0,y)]	\right.	\\
 &\left.+6y(1-y)  [G(0,r,y)-2 G(1,r,y)+ G(0,0,y)-2 G(1,0,y)]	\right.	\\
 &\left.-6 y^2(y+1)  G(0,0,0,y)-y(3 \mathrm{\pi} ^2 y^2+12 y^2+\mathrm{\pi} ^2 y-18 y+2 \mathrm{\pi} ^2+6) G(0,y)	\right.	\\
 &\left.-12 (1-y) (2 y^2+y+2) G(1,0,0,y)-6 (y+3) (y^2+1) \zeta (3)\right.	\\
 &\left.+(1-y)(12 y^2-\mathrm{\pi} ^2 y-24 y+12)\right]
 .
 \end{split}
\end{equation}

We see that, although $\mathcal{A}_0$ is the finite part of a 2-loop form factor, the highest weight of the GPLs that appears in Eq.~(\ref{FF}) is three. This happens because none of the master integrals that have $1/\epsilon^4$ poles contribute to the $gg \to H$ amplitude at leading order in the $\epsilon \to 0$ limit.

The expression for $\mathcal{F}$ in Eq.~(\ref{FF}) has been compared with a previous calculation in Ref.~\cite{Aglietti:2004nj} and agreement was found. The terms $\mathcal{A}_1(y)$ and $\mathcal{A}_2(y)$ are new. They can be found in the ancillary file provided with this paper.

\section{Conclusions}
\setcounter{equation}{0} 
\numberwithin{equation}{section}
We have presented a calculation of  the mixed two-loop QCD-electroweak 
corrections mediated by massless quarks 
to the production of the Higgs boson in gluon fusion. We extended 
the known result for these corrections to two higher orders in the dimensional 
regularization parameter $\epsilon$.  This is one of the  ingredients required for 
the computation of the NLO mixed QCD-electroweak corrections to $gg \to H$ amplitudes. 
We employed the method of differential equations, determined 
a canonical basis of master integrals and expressed all the 
relevant functions in terms of Goncharov polylogarithms.
Finally, we used a mixed numerical and analytical approach, based on the PSLQ algorithm,
in order to fix all necessary boundary conditions.
This establishes the necessary framework to successfully address the 
calculation of the missing three-loop virtual contributions, whose calculation is ongoing.

~

\section*{Acknowledgments}
The work of M.B. was supported by a graduate fellowship 
from Karlsruhe Graduate School ``Collider Physics at the highest energies 
and at the highest precision''. We wish to thank Christopher Wever for many fruitful discussions and help during all steps of this project, and Roberto Bonciani for his assistance with comparing our results with Ref.~\cite{Aglietti:2004nj}.

\begin{appendix}
\section{$\mathcal{F}_0$ values}
\label{AB}\setcounter{equation}{0} 
\numberwithin{equation}{section}
In this appendix we present the boundary conditions for the master integrals defined in Eq.~(\ref{DY}). The weights 0, 1 and 2 were determined analytically. For weights 4, 5 and 6 the results were obtained by fitting numerical results to an analytic \textit{Ansatz} to at least 750 digits.

\begin{equation}
 \mathcal{F}_1(y)=\epsilon^2 (\epsilon-1)(-s)^{2\epsilon} \imineq{0bubble}{7}
\end{equation}
\begin{equation}
\begin{split}
 \mathcal{F}_{0,1}^{(0)}=-\frac{1}{2}, \quad
 \mathcal{F}_{0,1}^{(1)}=0, \quad
 \mathcal{F}_{0,1}^{(2)}=-\frac{\pi^2}{6}, \quad
 \mathcal{F}_{0,1}^{(3)}=\zeta(3), \quad\\
 \mathcal{F}_{0,1}^{(4)}=-\frac{\pi^4}{20}, \quad
 \mathcal{F}_{0,1}^{(5)}=\frac{1}{3}\left[\pi^2\zeta(3)+9\zeta(5)\right], \quad
 \mathcal{F}_{0,1}^{(6)}=-\zeta^2(3)-\frac{61 \pi ^6}{3780}.
\end{split}
\end{equation}

~

~

\begin{equation}
 \mathcal{F}_2(y)=-\epsilon ^2 (-s)^{2\epsilon+1} \left[(\omega+1)\imineq{sbubble1}{7}+2\omega\imineq{sbubble2}{7} \right]
\end{equation}
\begin{equation}
\begin{split}
 \mathcal{F}_{0,2}^{(0)}=1, \quad
 \mathcal{F}_{0,2}^{(1)}=0, \quad
 \mathcal{F}_{0,2}^{(2)}=-\frac{\pi ^2}{3}, \quad
 \mathcal{F}_{0,2}^{(3)}=-10 \zeta (3), \quad\\
 \mathcal{F}_{0,2}^{(4)}=-\frac{11\pi ^4}{90}, \quad
 \mathcal{F}_{0,2}^{(5)}=\frac{10 \pi ^2 \zeta (3)}{3}-54 \zeta (5), \quad
 \mathcal{F}_{0,2}^{(6)}=50 \zeta^2(3)-\frac{121 \pi ^6}{1890}.
\end{split}
\end{equation}

~

~

\begin{equation}
 \mathcal{F}_3(y)=\epsilon ^2 (-s)^{2\epsilon+1} \imineq{sbubble2}{7}
\end{equation}
\begin{equation}
\begin{split}
 \mathcal{F}_{0,3}^{(0)}=0, \quad
 \mathcal{F}_{0,3}^{(1)}=0, \quad
 \mathcal{F}_{0,3}^{(2)}=\frac{\pi ^2}{6}, \quad
 \mathcal{F}_{0,3}^{(3)}=8 \zeta (3), \quad\\
 \mathcal{F}_{0,3}^{(4)}=\frac{7 \pi ^4}{72}, \quad
 \mathcal{F}_{0,3}^{(5)}=48 \zeta (5)-3 \pi ^2 \zeta (3), \quad
 \mathcal{F}_{0,3}^{(6)}=\frac{127 \pi ^6}{2160}-48 \zeta^2(3).
\end{split}
\end{equation}

~

~

\begin{equation}
 \mathcal{F}_4(y)=\epsilon^3 (-s)^{2\epsilon+1} \imineq{Triangle1}{7}
\end{equation}
\begin{equation}
\begin{split}
 \mathcal{F}_{0,4}^{(0)}=0, \quad
 \mathcal{F}_{0,4}^{(1)}=0, \quad
 \mathcal{F}_{0,4}^{(2)}=0, \quad
 \mathcal{F}_{0,4}^{(3)}=-2 \zeta (3), \quad\\
 \mathcal{F}_{0,4}^{(4)}=-\frac{\pi ^4}{180}, \quad
 \mathcal{F}_{0,4}^{(5)}=-\frac{\pi ^2 \zeta (3)}{3}-12 \zeta (5), \quad
 \mathcal{F}_{0,4}^{(6)}=9 \zeta^2(3)-\frac{37 \pi ^6}{3780}.
\end{split}
\end{equation}

~

~

\begin{multline}
 \mathcal{F}_5(y)=\epsilon^2 \left[-2(\epsilon-1) (-s)^{2\epsilon}\imineq{0bubble}{7}+\epsilon\frac{3(1+2\omega)}{\sqrt{1+4\omega}}(-s)^{2\epsilon+1} \imineq{Triangle1}{7}+\right.\\\left.-(-s)^{2\epsilon+2}\frac{\omega^2}{\sqrt{1+4\omega}}\imineq{Triangle2}{7}\right]
\end{multline}
\begin{equation}
\begin{split}
 \mathcal{F}_{0,5}^{(0)}=0, \quad
 \mathcal{F}_{0,5}^{(1)}=0, \quad
 \mathcal{F}_{0,5}^{(2)}=-\frac{\pi ^2}{3}, \quad
 \mathcal{F}_{0,5}^{(3)}=-4 \zeta (3), \quad\\
 \mathcal{F}_{0,5}^{(4)}=-\frac{41 \pi ^4}{180}, \quad
 \mathcal{F}_{0,5}^{(5)}=\pi ^2 \zeta (3)-30 \zeta (5), \quad
 \mathcal{F}_{0,5}^{(6)}=21 \zeta^2(3)-\frac{97 \pi ^6}{756}.
\end{split}
\end{equation}

~

~

\begin{multline}
 \mathcal{F}_6(y)=\epsilon^2\left[(1-\epsilon)(-s)^{2\epsilon}\frac{\sqrt{1+4\omega}}{2} \imineq{0bubble}{7}+(-s)^{2\epsilon+1} \frac{(\omega+1)\sqrt{1+4\omega}}{4}\imineq{sbubble1}{7}+\right.\\
 \left.+(-s)^{2\epsilon+1} \frac{(\omega+1)\sqrt{1+4\omega}}{2} \imineq{sbubble2}{7}+(-s)^{2\epsilon+2} \omega\sqrt{1+4\omega}\imineq{Inclusion}{7}\right]
\end{multline}
\begin{equation}
\begin{split}
 \mathcal{F}_{0,6}^{(0)}=0, \quad
 \mathcal{F}_{0,6}^{(1)}=0, \quad
 \mathcal{F}_{0,6}^{(2)}=\frac{\pi ^2}{4}, \quad
 \mathcal{F}_{0,6}^{(3)}=6 \zeta (3), \quad\\
 \mathcal{F}_{0,6}^{(4)}=\frac{5 \pi ^4}{48}, \quad
 \mathcal{F}_{0,6}^{(5)}=36 \zeta (5)-\frac{5 \pi ^2 \zeta (3)}{2}, \quad
 \mathcal{F}_{0,6}^{(6)}=\frac{77 \pi ^6}{1440}-36 \zeta^2(3).
\end{split}
\end{equation}

~

~

\begin{equation}
 \mathcal{F}_7(y)=\epsilon^4 (-s)^{2\epsilon+1} \imineq{DoubleTriangle}{7}
\end{equation}
\begin{equation}
\begin{split}
 \mathcal{F}_{0,7}^{(0)}=0, \quad
 \mathcal{F}_{0,7}^{(1)}=0, \quad
 \mathcal{F}_{0,7}^{(2)}=0, \quad
 \mathcal{F}_{0,7}^{(3)}=\zeta (3), \quad\\
 \mathcal{F}_{0,7}^{(4)}=\frac{5 \pi ^4}{72}, \quad
 \mathcal{F}_{0,7}^{(5)}=\frac{7 \pi ^2 \zeta (3)}{6}+6 \zeta (5), \quad
 \mathcal{F}_{0,7}^{(6)}=\frac{3 \zeta^2(3)}{2}+\frac{13 \pi ^6}{270}.
\end{split}
\end{equation}

~

~

\begin{equation}
 \mathcal{F}_8(y)=\epsilon^4 (-s)^{2\epsilon+1} \imineq{Arrow}{7}
\end{equation}
\begin{equation}
\begin{split}
 \mathcal{F}_{0,8}^{(0)}=0, \quad
 \mathcal{F}_{0,8}^{(1)}=0, \quad
 \mathcal{F}_{0,8}^{(2)}=-\frac{\pi ^2}{6}, \quad
 \mathcal{F}_{0,8}^{(3)}=4 \zeta (3), \quad\\
 \mathcal{F}_{0,8}^{(4)}=-\frac{\pi ^4}{9}, \quad
 \mathcal{F}_{0,8}^{(5)}=\frac{\pi ^2 \zeta (3)}{3}+20 \zeta (5), \quad
 \mathcal{F}_{0,8}^{(6)}=-16 \zeta^2(3)-\frac{173 \pi ^6}{3780}.
\end{split}
\end{equation}

~

~

\begin{equation}
 \mathcal{F}_9(y)=\epsilon^4 (-s)^{2\epsilon+2} \sqrt{4\omega+1} \imineq{NonPlanar}{7}
\end{equation}
\begin{equation}
\begin{split}
 \mathcal{F}_{0,9}^{(0)}=0, \quad
 \mathcal{F}_{0,9}^{(1)}=0, \quad
 \mathcal{F}_{0,9}^{(2)}=-\frac{\pi ^2}{3}, \quad
 \mathcal{F}_{0,9}^{(3)}=-24 \zeta (3), \quad\\
 \mathcal{F}_{0,9}^{(4)}=\frac{13 \pi ^4}{45}, \quad
 \mathcal{F}_{0,9}^{(5)}=\frac{46 \pi ^2 \zeta (3)}{3}-100 \zeta (5), \quad
 \mathcal{F}_{0,9}^{(6)}=264 \zeta^2(3)+\frac{397 \pi ^6}{945}.
\end{split}
\end{equation}

~

~

\begin{equation}
 \mathcal{F}_{10}(y)=\epsilon^2 (-s)^{2\epsilon+1} \imineq{btadpole}{7}
\end{equation}
\begin{equation}
\begin{split}
 \mathcal{F}_{0,10}^{(0)}=1, \quad
 \mathcal{F}_{0,10}^{(1)}=0, \quad
 \mathcal{F}_{0,10}^{(2)}=-\frac{\pi ^2}{6}, \quad
 \mathcal{F}_{0,10}^{(3)}=-2 \zeta (3), \quad\\
 \mathcal{F}_{0,10}^{(4)}=-\frac{\pi ^4}{40}, \quad
 \mathcal{F}_{0,10}^{(5)}=\frac{\pi ^2 \zeta (3)}{3}-6 \zeta (5), \quad
 \mathcal{F}_{0,10}^{(6)}=2 \zeta^2(3)-\frac{79 \pi ^6}{15120}.
\end{split}
\end{equation}

~

~

\begin{equation}
 \mathcal{F}_{11}(y)=-\epsilon^2 (-s)^{2\epsilon+2} \sqrt{4\omega+1} \imineq{doublebubble}{7}
\end{equation}
\begin{equation}
\begin{split}
 \mathcal{F}_{0,11}^{(0)}=0, \quad
 \mathcal{F}_{0,11}^{(1)}=0, \quad
 \mathcal{F}_{0,11}^{(2)}=\frac{\pi ^2}{6}, \quad
 \mathcal{F}_{0,11}^{(3)}=2 \zeta (3), \quad\\
 \mathcal{F}_{0,11}^{(4)}=-\frac{\pi ^4}{360}, \quad
 \mathcal{F}_{0,11}^{(5)}=6 \zeta (5)-\pi ^2 \zeta (3), \quad
 \mathcal{F}_{0,11}^{(6)}=-6 \zeta^2(3)-\frac{47 \pi ^6}{15120}.
\end{split}
\end{equation}

~

~

\begin{multline}
 \mathcal{F}_{12}(y)=\epsilon^2\left[\frac{1-\epsilon}{2}(-s)^{2\epsilon} \imineq{0bubble}{7}+ (-s)^{2\epsilon+1} \frac{\omega+1}{4} \imineq{sbubble1}{7}+(-s)^{2\epsilon+1} \frac{\omega+1}{2} \imineq{sbubble2}{7}+\right.\\
 \left.+(-s)^{2\epsilon+2} \omega \imineq{Inclusion}{7}+ (-s)^{2\epsilon+2} \imineq{doublebubble}{7}+(1-2\epsilon) (-s)^{2\epsilon+2} \omega \imineq{Kite}{7}\right]
\end{multline}
\begin{equation}
\begin{split}
 \mathcal{F}_{0,12}^{(0)}=0, \quad
 \mathcal{F}_{0,12}^{(1)}=0, \quad
 \mathcal{F}_{0,12}^{(2)}=\frac{\pi ^2}{12}, \quad
 \mathcal{F}_{0,12}^{(3)}=4 \zeta (3), \quad\\
 \mathcal{F}_{0,12}^{(4)}=\frac{77 \pi ^4}{720}, \quad
 \mathcal{F}_{0,12}^{(5)}=30 \zeta (5)-\frac{3 \pi ^2 \zeta (3)}{2}, \quad
 \mathcal{F}_{0,12}^{(6)}=\frac{1711 \pi ^6}{30240}-30 \zeta^2 (3).
\end{split}
\end{equation}

~

~

\end{appendix}

\bibliographystyle{BIBLIOSTYLE}
\bibliography{biblio.bib}

\end{document}